\documentclass[12pt,reqno,sumlimits]{amsart}
\usepackage{amssymb,pb-diagram}

\theoremstyle{plain}
\newtheorem{thm}{Theorem}

\newtheorem{prop}[thm]{Proposition}
\newtheorem{conj}[thm]{Conjecture}

\theoremstyle{definition}
\newtheorem{rem}[thm]{Remark}
\newtheorem{defn}[thm]{Definition}

\newcommand{\R}{{\mathbb R}}

\newcommand{\C}{{\mathbb C}}
\newcommand{\D}{{\mathbb D}}
\newcommand{\Z}{{\mathbb Z}}
\newcommand{\h}{{\mathbb H}}

\newcommand{\1}{{\bf 1}}

\newcommand{\g}{{\mathfrak  g}}
\newcommand{\uu}{{\mathfrak  u}}
\renewcommand{\h}{{\mathfrak  h}}
\newcommand{\ex}[1]{{e^{#1}}}
\newcommand{\calA}{{\mathcal A}}
\newcommand{\calB}{{\mathcal B}}

\newcommand{\calD}{{\mathcal D}}
\newcommand{\calE}{{\mathcal E}}

\newcommand{\calG}{{\mathcal G}}

\newcommand{\calI}{{\mathcal I}}

\newcommand{\calM}{{\mathcal M}}
\newcommand{\calO}{{\mathcal O}}
\newcommand{\calP}{{\mathcal P}}
\newcommand{\calR}{{\mathcal R}}

\newcommand{\calL}{{\mathcal L}}

\newcommand{\calU}{{\mathcal U}}
\newcommand{\calV}{{\mathcal V}}
\newcommand{\calZ}{{\mathcal Z}}
\renewcommand{\to}{\longrightarrow}

\newcommand{\iso}{\stackrel{\simeq}{\longrightarrow}}
\renewcommand{\iff}{\text{if and only if }}
\newcommand{\spinc}{$\text{spin}^c$}
\newcommand{\Dirac}{\not\!\!D}
\newcommand{\inc}{\hookrightarrow}

\newcommand{\End}{\operatorname{End}}
\newcommand{\Map}{\operatorname{Map}}

\newcommand{\Tr}{\operatorname{Tr}}
\newcommand{\Lie}{\operatorname{Lie}}
\newcommand{\ad}{{\operatorname{ad\,}}}

\newcommand{\rk}{{\operatorname{rk}}}

\newcommand{\id}{{\operatorname{id}}}
\newcommand{\ind}{{\operatorname{ind}}}
\newcommand{\Ind}{{\operatorname{Ind}}}

\newcommand{\cl}{{\operatorname{cl}}}

\newcommand{\inv}{^{-1}}

\newsavebox{\savepar}

\newcommand{\ip}[1]{\langle #1 \rangle}
\newcommand{\norm}[1]{\| #1 \|}

\numberwithin{equation}{section}

\begin{document}

\title[A Generalization of Witten's Conjecture]{A Generalization of Witten's
  Conjecture relating Donaldson and Seiberg--Witten Invariants}
\author{Adrian Vajiac}
\address{Department of Mathematics\\
  University of Texas at Austin}
\email{avajiac@math.utexas.edu}
\begin{abstract}
  We generalize Witten's conjectured formula relating Donaldson and
  Seiberg--Witten invariants to manifolds of non-simple type, via equivariant
  localization techniques. This approach does not use the theory of
  non-abelian monopoles, but works directly on the Donaldson--Witten and
  Seiberg--Witten moduli spaces.  We give a formal derivation of Witten's
  conjecture and its generalization, making use of an infinite dimensional
  version of the abelian localization theorem.
\end{abstract}

\maketitle



\section{Introduction}
\label{intro}

Topological quantum field theories (TQFT) emerged in the late 1980s as part
of the renewed relationship between differential geometry/topology and
physics.  In the 1990s, developments in TQFT gave unexpected results in
differential topology and symplectic and algebraic geometry.

One striking feature of physicists' approach to TQFT is the use of
mathematically non-rigorous Feynman path integrals to produce new topological
invariants of manifolds, which appear as the physical observables of the
TQFT.  For example, the first TQFT, formulated by Witten~\cite{ewdon} in 1988
(Donaldson--Witten theory), gives a quantum field theory representation of
Donaldson invariants.  Witten formulated two additional TQFT: 2D topological
sigma models~\cite{ewsig}, and 3D Chern-Simons gauge theory~\cite{ewdon}.
These theories are related, respectively, to Gromov--Witten invariants, and
to knot and link invariants (Jones polynomials and generalizations).  Seiberg
and Witten~\cite{nsew} extended these ideas to introduce the Seiberg--Witten
invariants of 4--manifolds. These invariants have revealed many deep
properties of 3-- and 4--manifolds and of symplectic manifolds.  Witten's
papers have stimulated a tremendous amount of mathematical research to make
his work mathematically rigorous (e.g. Axelrod and Singer~\cite{axsinger},
Kronheimer and Mrowka~\cite{km1}, Ruan and Tian~\cite{ruantian},
Taubes~\cite{taubes}).

This paper makes use of the Mathai--Quillen formalism~\cite{mq} to study
properties of Donaldson--Witten and Seiberg--Witten theories, and relations
between them. Basically, the Mathai--Quillen formalism can be applied to any
TQFT, with topological invariants defined by functional integrals of a
certain Euler class (or wedge products of the Euler class with other forms).
This approach was first applied by Atiyah and Jeffrey~\cite{aj} to Donaldson
theory and the Casson invariant, and was further developed in
e.g.~\cite{radu},\cite{wu}.  This approach is only formal at present, because
ordinary finite dimensional geometric constructions are applied to infinite
dimensional path integrals.  While a mathematically rigorous understanding of
path integrals is still unknown, their predictive power is well known in
physics, and their predictions in mathematics are equally striking.

In this paper, the Mathai--Quillen formalism for Donaldson--Witten and
Seiberg--Witten theories is stated in an equivariant setup, following the
work of Constantinescu~\cite{radu}, and the relevant invariants are obtained
from a functional integral point of view. Our main result is a generalization
of Witten's formula relating Donaldson and Seiberg--Witten invariants. We
state the main result as a conjecture, since the argument makes use of an
infinite dimensional version of the abelian localization theorem.

In more detail, Witten~\cite{ewmono} conjectured on physical grounds that
Donaldson--Witten invariants can be written in terms of Seiberg--Witten
invariants. This conjecture states the following.  Consider a compact,
connected, simply-connected, smooth 4--manifold $X$ with $b_2^+>1$.  Let
$\calA^*_E$ be the space of irreducible connections on an $SU(2)$--bundle $E$
over $X$, with second Chern class $c_2(E)=k$, and let $\calG_E$ be the group
of unitary gauge transformations of $E$. Then, if $\mu(\Sigma)\in
H^2(\calA^*_E/\calG_E)$ and $\mu(\nu)\in H^4(\calA^*_E/\calG_E)$ are the
so-called $\mu$--classes, the total Donaldson polynomial is given by:
$$
\D_X(\Sigma_1^{\alpha_1},\dots,\Sigma_{b_2}^{\alpha_{b_2}},\nu^\beta)
=\sum_{k\in\Z}\int_{\calM_k}\mu(\Sigma_1)^{\alpha_1}\dots
\mu(\Sigma_{b_2})^{\alpha_{b_2}}\mu(\nu)^\beta,
$$
where $\calM_k$ is the (Uhlenbeck compactification of the) moduli space of
Donaldson theory, consisting basically of anti-self-dual connections modulo
the gauge group.  Following Witten's notation, let us introduce formal
variables $q_1,\dots,q_{b_2},\lambda$, and write the generating function for
the total Donaldson polynomial as
\begin{equation*}
  \D_X\left(e^{\sum q_a\Sigma_a+\lambda\nu}\right)=
  \sum \frac{\D_X((q_1\Sigma_1)^{\alpha_1},\dots,(q_{b_2}
    \Sigma_{b_2})^{\alpha_{b_2}},(\lambda\nu)^\beta)}{\alpha_1!\cdots
    \alpha_{b_2}!\cdot\beta!}.
\end{equation*}
Let $SW(x)$ denote the Seiberg--Witten invariant for each isomorphism class
$x$ of \spinc--structures on $X$, which counts the number of points in the
zero dimensional SW moduli space.  For $v=\displaystyle\sum_a q_a\Sigma_a$, the
conjectured equality for manifolds of simple type is~\cite{ewmono}
\begin{eqnarray}
  \label{conjintro}
  \D_X(\ex{\sum_a q_a\Sigma_a + \lambda\nu})&=&\nonumber\\
  2^{\scriptstyle{1+\frac{1}{4}(7\chi+11\sigma)}}
  &\cdot&\left[\ex{\left(\frac{v^2}{2}+2\lambda\right)}
    \sum_x SW(x)\cdot\ex{\scriptstyle{v\cdot x}}+i^{\frac{\chi+\sigma}{4}}
    \ex{\left(-\frac{v^2}{2}-2\lambda\right)}
    \sum_x SW(x)\cdot\ex{\scriptstyle{-iv\cdot x}}\right],\nonumber\\
\end{eqnarray}
where $\chi,\sigma$ denote the Euler characteristic and signature of $X$,
respectively, and the sums run over all basic classes $x$.

While this conjecture checks in all known examples, Witten's
derivation~\cite{ewmono} of the formula uses physical arguments whose
mathematical content is unclear at present (see~\cite{radu},~\cite{park} for
other physical derivations). In a series of papers~\cite{feehan}, Feehan and
Leness propose a rigorous derivation of this conjecture, following the work
of~\cite{pt}.  These authors have made excellent progress towards the
conjecture, although the proofs are quite different from Witten's original
intuition.  In particular, the authors work on the moduli space of
non-abelian monopoles.

This paper takes a different approach towards Witten's conjecture.  Some
steps of this argument are still formal at present.  In particular,
following~\cite{radu}, equivariant localization techniques are used to reduce
Witten's conjecture to computations on the fixed point set of a group action.
This is closer in spirit to recent proofs~\cite{givental},~\cite{lly} of the
mirror symmetry conjecture.  In general, it seems that equivariant cohomology
is a more natural domain for TQFT observables than the usual cohomology
groups, since the configuration spaces of TQFT come with group actions.

In contrast of working on the moduli spaces of non-abelian monopoles, we
introduce group actions on the moduli spaces of Seiberg-Witten solutions
(abelian monopoles) and anti-self-dual connections.  The relevant equivariant
path integrals can be compared by localization techniques, since the fixed
point sets of the group actions coincide, even though the total moduli spaces
differ.

The main result of this paper is a derivation of a generalization of Witten's
formula to equivariant invariants and manifolds of non-simple type, via
equivariant localization techniques. Let
$\D_X\left(e^{\Sigma+\lambda\nu}\right)(m)$ be the equivariant Donaldson
generating series, and $SW(c,m)$ be the equivariant SW invariant for a
\spinc-structure $c$, as defined in $\S$\ref{dwmq}, and $\S$\ref{swmq},
respectively. Let $2s(c)$ be the dimension of the SW moduli space for the
corresponding \spinc-structure $c$. We obtain the following conjectured
formula relating equivariant Donaldson and Seiberg--Witten invariants,
equality which holds in $\C[m]$, the $S^1$-equivariant cohomology ring of a
point:
\begin{eqnarray}
  \label{Wgenformula}
  &&\D_X\left(e^{\Sigma+\lambda\nu}\right)(m) = 2^{1+\frac{7\chi+11\sigma}{4}}
  \sum_c 2^{8s(c)}\left(\frac{2\pi}{m}\right)^{2s(c)}
  SW(c,m)\cdot e^{\left(\frac{m}{2\pi}\right)(c\cdot\Sigma+\frac{1}{2}\Sigma^2)}
  \cdot e^{\left(\frac{m}{2\pi}\right)^2 (2\lambda)}\nonumber\\
  &&\qquad +2^{1+\frac{7\chi+11\sigma}{4}} i^{\frac{\chi+\sigma}{4}}
  \sum_c 2^{8s(c)}\left(\frac{2\pi}{m}\right)^{2s(c)}
  SW(c,m)\cdot e^{\left(\frac{m}{2\pi}\right)(-ic\cdot\Sigma-\frac{1}{2}\Sigma^2)}
  \cdot e^{\left(\frac{m}{2\pi}\right)^2 (-2\lambda)}.\nonumber\\
  \,
\end{eqnarray}
\eqref{Wgenformula} reduces to Witten's formula~\eqref{conjintro},
identifying the degree zero components (the expression being regular at
$m=0$), and assuming that $X$ has simple type.  This derivation assumes that
finite dimensional equivariant localization formulas extend to infinite
dimensions, so the argument has a formal part.  However, it is possible that
the equivariant path integrals used can be made rigorous, which would provide
an alternative to the unsuccessful search for good measures on infinite
dimensional spaces of connections.

Since this paper covers topics in both mathematics and physics, we have
assembled material from both fields (some well-known and some quite
specialized) in the initial sections.  Section~\ref{mq} covers standard
mathematical material on Mathai--Quillen formalism, specialized results on
extensions to the equivariant setting (following the work of~\cite{radu}),
and a discussion of the physicists' version of Mathai--Quillen formalism in
infinite dimensions.  Section~\ref{mq_tqft} discusses the application of the
Mathai--Quillen construction to the infinite dimensional settings for
Donaldson--Witten and Seiberg--Witten theories, in the spirit of~\cite{radu}.
Section~\ref{equivloc} is the main part of this paper, where abelian
localization techniques are applied to the setups of section~\ref{mq_tqft},
and new descriptions of Donaldson polynomial and Seiberg--Witten invariants
are obtained.  Section~\ref{conjecture} is devoted to a complete derivation
our generalization of Witten's conjecture, and a derivation of Witten's
formula~\eqref{conjintro} from~\eqref{Wgenformula}.


\section{The Mathai-Quillen Formalism}
\label{mq}

The Mathai--Quillen (MQ) formalism~\cite{mq} can be applied to any TQFT. This
approach was first implemented by Atiyah and Jeffrey~\cite{aj}, and later
further developed in~\cite{blau},~\cite{radu},~\cite{cordes}.  The basic idea
behind this formalism is the extension to the infinite-dimensional case of
ordinary finite-dimensional geometric constructions.  MQ formalism gives a
unified description of many Cohomological Field Theories (CFT) and
Supersymmetric Quantum Mechanics (SQM), and it also provides some insight
into the mechanism of the localization of path integrals in SQM and TQFT.

\subsection{The Mathai--Quillen Construction}
\label{mqfinitecase}

Let us start with a summary of definitions and properties of equivariant
cohomology, which will be used later. For details one should
see~\cite{ab},~\cite{bgv},~\cite{radu},~\cite{cordes}.

Let $M$ be a manifold, and $G$ a connected compact Lie group with Lie algebra
$\g$, such that $G$ acts on $M$. The equivariant cohomology $H^*_G(M)$ is
defined to be the ordinary cohomology $H^*(M^G)$ of the space $M^G=EG\times_G
M,$ where $EG$ is the universal bundle over $BG$, the classifying space of
$G$.  Equivariant cohomology is a contravariant functor from $G$-spaces to
modules over the base ring $H^*_G(\text{pt})=H^*(BG).$

If $H\subset G$ is a closed subgroup, then $H^*_G(G/H)=H^*_H(\text{pt})$.
Consider the inclusion $i:M\inc M^G$, via the fiber over a base point of
$BG$, which induces a natural homomorphism from equivariant cohomology to
ordinary cohomology
$$
i^*:H^*_G(M)\to H^*(M).
$$
For $T$ a maximal torus in $G$, we have
\begin{equation}
  \label{maxtorus}
  H^*_G(M)\iso H^*_T(M)^W,
\end{equation}
where the right hand side consists of elements of $H^*_T(M)$ which are
invariant under the Weyl group $W$.  \eqref{maxtorus} reflects the
fact that the study of compact Lie groups, which reduces to the behavior of
the Weyl group acting on a maximal torus of $G$, also applies to the
equivariant theory~\cite[p. 4]{ab}.

There are also algebraic models for defining the equivariant cohomology. The
most common are the Weil model and the Cartan model. Although they are
equivalent (see e.g.~\cite{bgv},~\cite{cordes}), the Cartan model is more
natural in TQFT. We summarize it below.  Consider the graded algebra of
equivariant differential forms
$$
\Omega_G(M)=\left[S(\g^*)\otimes\Omega(M)\right]^G\otimes\C ,
$$
where the $\Z$-grading is the usual on $\Omega(M)$, and twice the grading
on $S(\g^*)$, $[\,\cdot\,]^G$ is the subalgebra of $G$-invariant elements,
and the action of $G$ on $\Omega(M)$ is obtained from the action of $G$ as
diffeomorphisms on $M$, and on $S(\g^*)$ is obtained from symmetric powers of
the coadjoint action.  The equivariant  deRham graded differential of degree
$+1$ on $\Omega_G(M)$ is defined by
$$
d_G\alpha(X)=d[\alpha(X)] - \iota_{\rho(X)}[\alpha(X)] ,
$$
where we extend $\alpha$ to be in $S^i(\g^*)\otimes\Omega^j(M)$, so
$d\alpha\in S^i(\g^*)\otimes\Omega^{j+1}(M)$. Moreover,
$\iota_{\rho(\cdot)}\in S^{i+1}(\g^*)\otimes\Omega^{j-1}(M)$.  Here $d$
denotes the usual deRham differential, $\iota$ is the interior product, and
$\rho:\g\to\hbox{ Vect}(TM)$ is the infinitesimal action of $\g$.  One can
show that $d_G^2=0$. Then the $G$-equivariant cohomology of $M$ with complex
coefficients is defined as the cohomology of the complex above, and this
cohomology equals $H^*_G(M;\C)$.

If $\alpha$ is equivariantly closed for all $X\in\g$, i.e.  $d_G\alpha=0$,
and if we decompose $\alpha$ into its homogenous components, then
$$
\iota_{\rho(X)}\alpha(X)_{[i]}=d\alpha(X)_{[i-2]}.
$$
Moreover, one can show that $H^*_G(\text{pt})=S(\g)^G\otimes\C$. In
particular, for $G=S^1$, $H^*_{S^1}(\text{pt})=\C[m].$ Then integration of
$G$-equivariant forms, $\displaystyle\int:\Omega^*_G(M)\to S(\g)^G, $ is
given by the Berezin integral, as defined in Definition~\ref{berezin}.

For $G=S^1$, the coadjoint action is trivial, so
$$
\Omega_{S^1}(M)=\C[m]\otimes\Omega(M)^{S^1},
$$
where $m$ is a generator of $\uu(1)^*$, the dual of the Lie algebra of
$S^1$, and
$$
\Omega(M)^{S^1}=\left\{\omega\in\Omega(M)|\calL_X\omega=0\right\},
$$
where
$X$ is the vector field corresponding to the generator of the Lie algebra
of $S^1$, dual to $m$.  Then if $\omega\in\Omega^{S^1}(M)$ and $k\in\Z$,
$$
d_{S^1}(m^k\omega)=m^k d\omega - m^{k+1}\iota_X\omega.
$$

\begin{rem}
  The BRST operator of a TQFT, as defined in the physics literature, can be
  interpreted as the differential $d_G$ for a $G$-equivariant cohomology of a
  certain space of fields. Note that in some models $G$ might be trivial.
\end{rem}

The Mathai--Quillen (MQ) formalism was first defined in~\cite{mq} and then
refined in a series of later papers and books (e.g.~\cite{aj}).  The setup
for the MQ construction is the following (the organization of the material
and the notation below is based on~\cite{radu}).  Let $M$ be a compact
oriented manifold, $G$ a compact Lie group, $P$ a principal $G$-bundle over
$M$, $V$ an oriented inner product space of $\dim{V}=2m$, and $\rho_V:G\to
SO(V)$ a representation of $G$.  Let $E=P\times_{\rho_V}V$ be the associated
vector bundle.  One can identify $P$ with the bundle of all orthonormal
oriented frames on $E$.  The MQ construction produces {\em universal Thom
  forms} $U_t\in\Omega_G(V)$ for $t\in\R_+.$
\begin{defn}
  \label{berezin}
  Consider $A$ a commutative superalgebra, and $A\otimes\Lambda(V)$ the
  graded tensor product.  Choose a (non-zero) volume element
  $\text{vol}\in\Lambda^{\text{top}}(V)$. Then the {\em Berezin integral} is
  the map
  $$
  \calB_\chi:A\otimes\Lambda(V)\to A
  $$
  given by $\calB_\chi(a)=$ the coefficient of $a^{\text{top}}$, where
  $a^{\text{top}}\in A\otimes\Lambda^{\text{top}}(V)\simeq
  A\otimes\R\cdot\text{vol}$.
\end{defn}
In what will follow we will consider the case $A=S(\g^*)\otimes\Lambda(V)$.
\begin{defn}
  For all $t\in\R_+$, the {\em universal Thom form} 
  $U_t\in S(\g^*)\otimes\Lambda(V)$ is defined by
  $$
  U_t=(2\pi t)^{-m}\calB_\chi\left(\ex{\frac{t}{2}(-\norm{x}^2
      -2i\sum_i dx_i\cdot\chi_i + \sum_{i,j}\chi_i\cdot\phi_V\chi_j)}\right)
  $$
  where the notation is explained as follows:
  \begin{itemize}
  \item[a)] $x_i$ are coordinates on $V$ dual to $\chi_i$.
  \item[b)] $\phi_V\in S^1(\g^*)\otimes so(V)$ is obtained as follows.
    Let
    $$
    \phi\in S^1(\g^*)\otimes\g\simeq\g^*\otimes\g\simeq\End(\g),
    $$
    be the universal Weil element corresponding to the identity.  Apply
    $d\rho:\g\to\End(V)$ to the $\g$ part of $\phi$, and call the result
    $\phi_V$.  We may regard $\phi_V$ as the $G$-universal curvature matrix
    with entries $(\phi_V)_{jk}$. Here $\phi_V$ acts on the fibers of
    $E$.
  \item[c)] $\chi_i\cdot\phi_V\otimes\chi_j\in
    S^1(\g^*)\otimes\Lambda(V)\otimes\Lambda(V)$, where $\cdot$ denotes the
    exterior product in $\Lambda(V)$.
  \item[d)] $\displaystyle\sum_i dx_i\cdot\chi_i=\displaystyle\sum_i
    dx_i\otimes\1\otimes\chi_i\in S(\g^*)\otimes\Lambda(V)\otimes\Lambda(V)$.
  \item[e)] The exponential factor in the definition above lies in
    $S(\g^*)\otimes\Lambda(V)\otimes\Lambda(V)$, so its Berezin integral is
    in $S(\g^*)\otimes\Omega(V).$
  \end{itemize}
\end{defn}
We can use the shorthand notation
$$
U_t=(2\pi t)^{-m}\ex{-\frac{tx^2}{2}}
\calB_\chi\left(\ex{\frac{t}{2}(-2idx\chi + \chi\phi\chi)}\right),
$$
where we regard $x,dx,\chi$ as vectors and $\phi$ as a matrix.  The factor
$\ex{-\frac{tx^2}{2}}$ guarantees the rapid decay of $U_t$ as differential
forms on $V$.

The Thom forms $U_t$ are $G$-invariant (i.e.  $U_t\in\Omega_G(V)$, $d_G
U_t=0$), and $\displaystyle\int_V U_t=1$ (see~\cite[p. 26-28]{radu}).
Moreover, for all $t\in\R_+$ and sections $s$, $U_t(s)=s^*U_t$ represents the
cohomology class in $H^{2m}_G(V)$ corresponding to $\1\in H^0_G(\text{pt})$.
Here we view $V$ as a $G$-equivariant bundle over a point.

Via the universal Thom forms $U_t$, the Thom forms on $E$ are constructed as
follows. The horizontal projection composed with the Chern-Weil map~\cite[p.
15]{radu}
$$
\Omega_G(V)\stackrel{CW}{\to}\Omega(P\times V)^G
\stackrel{\text{Hor}}{\to}\Omega(E)
$$
is a chain map with respect to $d_G$ on $\Omega_G(V)$ and $d$ on
$\Omega(E)$.  Thus we define the Thom forms on $E$ by
$$
Th_t(E)=\text{Hor}\circ CW(U_t),
$$
One can prove that $Th_t(E)\in\Omega^{2m}(E)$, $d Th_t(E)=0$, and
$\displaystyle\int_{E_x} Th_t(E)=1$.

The Euler classes $e_t(E,s)$ are obtained form a section $\hat{s}:M\to E$ (or
equivalently, from a $G$-equivariant map $s:P\to V$), using the following
commutative diagram:
\[
\begin{diagram}
  \node{U_t\in\Omega_G(V)}\arrow{e,t}{\text{Hor }\circ\, CW}\arrow{s,l}{s^*}
  \node{\Omega(E)\ni Th_t(E)}\arrow{s,r}{\hat{s}^*}\\
  \node{U_t(s)\in\Omega_G(P)}\arrow{e,t}{\text{Hor }\circ\, CW}
  \node{\Omega(M)\ni e_t(E,s)}
\end{diagram}
\]
In the diagram above
$$
U_t(s)=(2\pi t)^{-m}\calB_\chi\left( \ex{\frac{t}{2}(-\norm{s}^2-2ids\chi
    + \chi\phi\chi)}\right).
$$
If $\dim{V}=\dim{M}$, then the Euler number is given by
$$
\chi(E)=\int_M e_t(E,s),
$$
for all $t$ and $s$. Moreover, one can show that
$$
\chi(E) = (2\pi t)^{-m}\int_M\text{Hor}\circ CW(\alpha),
$$
where $\alpha=(2\pi t)^{m}U_t(s)$ is an equivariantly closed form in
$\Omega_G(P)$, and therefore defines a class in $H^*_G(P)$.

Since the action of $G$ on $P$ is free, $\text{Hor}\circ
CW_{\theta}=(\pi^*)\inv,$ where $P\stackrel{\pi}{\to}P/G=M$.  Then
$$
\int_M\text{Hor}\circ CW_{\theta}(s^*(U_t)) = \int_M(\pi^*)\inv(s^*(U_t))
=\int_M\int_{TM}\calD\eta\, (\pi^*)\inv(s^*(U_t)),
$$
where $\eta$ is a fermionic variable on $TM$ which integrates to one.  Let
us write $\Omega=d\theta+\frac{1}{2}[\theta,\theta]$. The horizontal part is
$$
\Omega_{\text{hor}}=\phi=\text{Hor}\left(d\theta+\frac{1}{2}[\theta,\theta]\right),
$$
where $\phi$ is defined as above. Moreover, one can show that
$\chi\phi\chi=\displaystyle\frac{1}{2}R^{ij}_{kl}\eta^k\eta^l\chi_i\chi_j$,
so one obtains the following expression for $\chi(E)$, appearing in physics
literature (see e.g.~\cite{labastida_marino1}):
$$
\chi(E)=(2\pi t)^{-m}\int_M\,dx\int_{TM}\,\calD\eta\,\calB_\chi\left(
  \ex{-\frac{t}{2}\norm{s(x)}^2 
    +\frac{t}{2}(-2i\nabla_i s^j(x)\eta^i\chi_j+
    \frac{1}{2}R^{ij}_{kl}\eta^k\eta^l\chi_i\chi_j)}\right).
$$
In the definition of $\chi(E)$, there is no dependence on $t$ or on the
section $s$. Then for $t\to\infty$ we obtain
$$
\chi(E)=\sum_{\text{zeros of }s}\pm 1,
$$
and for $t\to 0$, we recover the definition of the Euler number:
$$
\chi(E)=(2\pi)^{-m}\int\text{Pf}(\Omega),
$$
For details, one should see e.g.~\cite[p. 111]{cordes}.


\subsection{The Mathai-Quillen Construction for Equivariant Vector Bundles}
\label{mqeqvb}

In the framework of the MQ formalism, an equivariant extension of the Thom
form with respect to a vector field action, has been constructed by Labastida
and Mari\~no~\cite{labastida_marino2}. In this paper, the authors analyze in
detail the cases of topological sigma models and non-abelian monopoles on
four-dimensional manifolds.  In a similar context, we summarize below an
extension of the MQ construction to equivariant vector bundles,
following~\cite[Secs. 2.3, 2.6]{radu}.

As before, let $M$ be a compact oriented manifold.  Let $H$ be a subgroup of
$G$, and $P$ be a $H$-equivariant $G$-bundle over $M$, i.e. $H$ acts on $P$
and $M$, and the action of $H$ on $P$ commutes with the action of $G$ on $P$.
Here $G$ and $H$ are compact Lie groups, with $H$ connected (such that the
action on $E$ below preserves the orientation).  Let $V$ be a
$2m$-dimensional oriented vector space with inner product and
representations:
$$
\rho:G\to SO(V),\qquad\lambda:H\to SO(V).
$$
Set $E=P\times_G V$, where the action of $H$ on $E$ is induced by the
action of $G$, so $E$ is naturally an $H$-equivariant bundle over $M$.

One needs to replace the Chern-Weil map $CW$ by an equivariant version. If we
take an $H$-invariant connection on $P$, and consider its $H$-equivariant
1-form $\theta\in\Omega^1(P)\otimes\g,$ i.e. $\calL_Y\theta=0$, for all
$Y\in\h=\Lie(H)$, then its curvature $\Omega\in\Omega^2(P)\otimes\g.$ Then
one can define the equivariant moment $\calI$ of the connection $\theta$
(analogous to the moment map in symplectic geometry), given by
$$
\calI(Y)=\iota_Y\theta\in C^\infty(P)\otimes\g,
$$
for all $Y\in\h$, where $\iota_Y$ is the vector field on $P$ determined by
$Y\in\h$ through the infinitesimal action of $\h$ on $P$.  Then the
$H$-equivariant curvature is given by
$$
\Omega_H=\Omega-\calI\in S(\g^*)\otimes\Omega(P)\otimes\g.
$$
One can show the following:
\begin{prop}
  For all $X\in\g$, we have
  $$
  \calL_X(\Omega_H)=-\text{ad}_X\circ\Omega_H.
  $$
  Since $\calL_Y\theta=0$, for all $Y\in\h$,
  $\theta\in\Omega^1_H(P)\otimes\g$, and 
  $$
  \Omega_H=d_H\theta+\frac{1}{2}[\theta,\theta]
  $$
  (the equivariant Maurer-Cartan equations).
  For $D$ the covariant derivative of $\theta$, we define $D_H$ on
  $\Omega_H(P)$ by:
  $$
  D_H\omega(Y)=D\omega(Y)-\iota_Y[\omega(Y)]
  $$
  for $Y\in\h$. Then $D_H\Omega_H=0$ (the equivariant Bianchi identity).
\end{prop}

The $H$-equivariant $CW$ map is the algebra homomorphism
$CW_{H,\theta}:S(\g^*)\to\Omega_H(P)$ defined by:
$$
CW_{H,\theta}(X^*)=X^*(\Omega_H)\in\Omega_H(P),
$$
for all $X^*\in\g^*$.  Note that $CW_{H,\theta}$ extends to a map
$CW_{H,\theta}:\Omega_{G\times H}(P)\to\Omega_H(P)^G,$ inducing the
$H$-equivariant $CW$ homomorphism:
$$
\text{Hor}\circ CW_{H,\theta}:\Omega_{G\times H}(P)\to\Omega_H(M).
$$
\begin{prop}~\cite[p. 31]{radu}
  The following hold:
  \begin{enumerate}
  \item $\text{Hor}\circ CW_{H,\theta}$ is a chain map with respect to
    $d_{G\times H}$ on $\Omega_{G\times H}(P)$, and $d_H$ on $\Omega_H(M)$,
    so $\text{Hor}\circ CW_{H,\theta}$ descends to a map
    $$
    H^*_{G\times H}(P)\to H^*_H(M).
    $$
  \item $\text{Hor}\circ CW_{H,\theta}:H^*_{G\times H}(P)\to H^*_H(M)$ is
    independent of $\theta$.
  \item $\text{Hor}\circ CW_{H,\theta}:H^*_{G\times H}(P)\iso H^*_H(M)$ is an
    isomorphism.  
  \end{enumerate}
\end{prop}
The MQ construction can now be extended to equivariant vector bundles. We define
as before a form $U_t\in\Omega_{G\times H}(V),$ and let
$Th_{H,t}(E)=\text{Hor}\circ CW_{H}(U_t)\in\Omega_H(E),$ where
$$
\Omega_{G\times H}(V)\stackrel{CW_H}{\to}\Omega_H(P\times V)^G
\stackrel{\text{Hor}}{\to}\Omega_H(E).
$$

Then $\deg(Th_{H,t})=2m$, $d_H Th_{H,t}=0$, and $\displaystyle\int_{E_x}Th_{H,t}=1.$

We have the following commutative diagram:
\[
\begin{diagram}
  \node{U_t\in\Omega_{G\times H}(V)}\arrow{e,t}{\text{Hor}\circ CW_H}
  \arrow{s,l}{s^*}
  \node{\Omega_H(E)\ni Th_{H,t}(E)}\arrow{s,r}{\hat{s}^*}\\
  \node{U_{H,t}(s)\in\Omega_{G\times H}(P)}\arrow{e,t}{\text{Hor}\circ CW_H}
  \node{\Omega_H(M)\ni e_{H,t}(E,s)}
\end{diagram}
\]
where $s:M\to E$ is an $H$-equivariant section.
More explicitly,
$$
U_{H,t}(s)=\calB_\chi\left((2\pi t)^{-m}\ex{\frac{t}{2}
    (-\norm{s}^2-2ids\chi+\chi(\phi_G+\phi_H)\chi)}\right).
$$


\subsection{Refinements}
\label{mqrefinements}

In this section we give an alternative version of the MQ integral, which
expresses the Euler number as an integration over $P$ instead of over $M$.
This section is based on~\cite[Sec. 2.4, 2.6]{radu}.

Choose a vertical volume element $v$ (volume element along the fibers of
$\pi:P\to M$) normalized such that its integral over the fibers is 1.  Then
for all $\beta\in\Omega^{\text{top}}$ we have
$$
\int_P \pi^*\beta\wedge v = \int_M \beta.
$$

\begin{defn}
  The {\em equivariant vertical volume element} $\gamma_G\in\Omega(P)\otimes
  S(\g)$ is constructed as follows.  Let $\lambda_1,\dots,\lambda_{\dim{\g}}$
  be an orthonormal basis of $\g$ with respect to an inner product (e.g. the
  Killing form for a semisimple Lie group), normalized such that
  $\text{vol}(G)=1.$ Choose a $G$-invariant connection $\theta$ on $P$ with
  curvature $\Omega$. If $\eta=(\eta_1,\dots,\eta_{\dim{\g}})$ is a second
  basis of $\g$, we set
  $$
  \gamma_G = \ex{\sum_\alpha\Omega_\alpha\otimes\lambda_\alpha}
  \calB_\eta\left(\ex{\sum_\alpha\theta_\alpha\otimes\eta_\alpha}\right).
  $$
\end{defn}

If we write $\theta=\theta_\alpha\otimes\lambda_\alpha,$ and
$\Omega=\Omega_\alpha\otimes\lambda_\alpha,$ then
$$
\ex{\sum_\alpha\Omega_\alpha\otimes\lambda_\alpha}\in \Omega(P)\otimes
S(\g),\qquad
\calB_\eta\left(\ex{\sum_\alpha\theta_\alpha\otimes\eta_\alpha}\right)\in
\Omega(P)\otimes\Lambda(\g),
$$
and we can write
$$
\gamma_G= \ex{\Omega\otimes\lambda}
\calB_\eta\left(\ex{\theta\otimes\eta}\right).
$$

\begin{prop}
  \label{propabove}
  The following hold:
  \begin{enumerate}
  \item The linear functional on $\Omega_G(P)$
    $$
    \alpha\mapsto\int_P\ip{\alpha\wedge\gamma_G}
    $$
    (contraction of the polynomial parts of $\alpha$ and $\gamma_G$)
    vanishes on $d_G$-exact forms, and hence descends to a map 
    $H^*_G(P)\to H^*_G(\text{pt})$.
  \item Let $\alpha^P\in H^*_G(P)$ and 
    $\alpha^M=\text{Hor}\circ CW(\alpha^P)\in H^*(M)$. Then
    $$
    \int_P\ip{\alpha^P\wedge\gamma_G} = \int_M\alpha^M.
    $$
  \end{enumerate}
\end{prop}
For a proof of this proposition, see~\cite{austin}.  Note that
$\alpha^P\in\Omega(P)\otimes S(\g^*)$, and $\gamma_G\in\Omega(P)\otimes
S(\g)$, so the pairing $\ip{\,,\,}$ is the pairing between $S(\g^*)$ and
$S(\g)$.  Moreover, if $\alpha^P=\pi^*\beta$, with
$\beta\in\Omega^{\text{top}}(M)$, then by a degree count, the only part of
$\gamma_G$ involved is the usual vertical volume element.  The pairing
between $\alpha$ and $\gamma_G$ substitutes the curvature in the polynomial
part of $\alpha$. The result is then just multiplied by the vertical volume
element.

Applying Proposition~\ref{propabove} to the MQ construction, one obtains:
\begin{eqnarray}
  \label{chiofE}
  \chi(E)&=&(2\pi t)^{-m}(2\pi)^{-\dim\g}\int_P\ex{-\frac{t}{2}\norm{s}^2}\nonumber\\
  &&\qquad\cdot\int_{\lambda\in\g}\int_{\phi\in\g}
  \ex{i\ip{\lambda,\phi}}\ex{-i\Omega\otimes\lambda}
  \calB_\chi\left(\ex{\frac{t}{2}(-2ids\chi+\chi\phi\chi)}\right).
\end{eqnarray}
In order to be rigorous, one should introduce a convergence factor
$\ex{-\epsilon\ip{\phi,\phi}}$, and take the limit when $\epsilon\to 0$.

More explicitly, let us choose a $G$-invariant metric on $P$, which induces a
connection $\theta$ with curvature $\Omega$, whose horizontal distribution is
given by the orthogonal complements to the tangent space to the $G$-orbits.
Now let $C:\g\to TP$ be the infinitesimal action of $\g$, and
$C^*\in\Omega^1(P)\otimes\g$ its adjoint with respect to the inner products
on $\g$ and $TP$. Then $\theta=(C^*C)\inv C^*$, and
$$
\Omega_{\text{hor}}=\text{Hor}\left(d\theta +
  \frac{1}{2}[\theta,\theta]\right)=(C^*C)\inv dC^*.
$$
In the formula~\eqref{chiofE}, $\Omega_{\text{hor}}$ can be used instead of
$\Omega$, since $\gamma_G$ is a top degree vertical form.  Finally one
obtains:
\begin{eqnarray}
  \label{chiofEfinal}
  \chi(E)&=&(2\pi t)^{-m}(2\pi)^{-\dim\g}\int_P\ex{-\frac{t}{2}\norm{s}^2}
  \calB_\eta\left(\ex{(C^*,\eta)}\right)\cdot\nonumber\\
  &&\qquad\cdot\int_{\lambda,\phi\in\g}
  \ex{i\ip{\phi,C^*C\lambda}}\ex{-idC^*\otimes\lambda}
  \calB_\chi\left(\ex{-itds\chi+\frac{t}{2}\chi\phi\chi)}\right)
\end{eqnarray}
We set:
$$
\Gamma_G=\calB_\eta\left(\ex{(C^*,\eta)}\right)
\ex{i\ip{\phi,C^*C\lambda}-idC^*\otimes\lambda}.
$$
Then~\eqref{chiofEfinal} can be written as follows:
\begin{equation}
  \label{chi}
  \chi(E)=(2\pi t)^{-m}(2\pi)^{-\dim\g}\int_P\ex{-\frac{t}{2}\norm{s}^2}
  \int_{\lambda,\phi\in\g}\Gamma_G\wedge
  \calB_\chi\left(\ex{-itds\chi+\frac{t}{2}\chi\phi\chi)}\right)
\end{equation}
Integration over $\lambda$ yields a delta function in $\phi$, centered at
$(C^*C)\inv dC^*=\Omega_{\text{hor}}$. This means that $\phi$ can be replaced
by $\Omega_{\text{hor}}$.

As computed by Constantinescu~\cite[Sec. 2.4, 2.6]{radu}, the
formula~\eqref{chi} can be refined by replacing $\Gamma_G$ with
$$
\Gamma_G(t)=\calB_\eta\left(\ex{t(C^*,\eta)}\right)
\ex{it\ip{\phi,C^*C\lambda}-itdC^*\otimes\lambda}
$$
which introduces the ``coupling constant'' $t$.  Changing the variables:
$\lambda\mapsto t\frac{\lambda}{2}$, $\eta\mapsto t\frac{\eta}{2},$ one
obtains:
\begin{eqnarray*}
  \chi(E)&=&(2\pi)^{-\dim G}(2\pi t)^{-m}\int_P\int_{\lambda,\phi\in\g}
  \calB_\eta\calB_\chi\left(e^{\frac{t}{2}(-\norm{s}^2 + (C^*,\eta) +
      i\ip{\phi,C^*C\lambda}-dC^*\otimes\lambda-2ids\chi+\chi\phi\chi)}\right)\\
  &=&(2\pi)^{-\dim G}\int_{P,\g}\Gamma_G(t)\wedge U_t(s).
\end{eqnarray*}
which agrees with the results of~\cite{aj}.

An important aspect, which will be crucial in $\S$\ref{equivloc}, is
that one can write the same refinements for groups $G$ which split as
$G_0\times S^1$, with the assumption that $G_0$ acts freely on $P$
(see~\cite[Sec 2.6]{radu}). Then $H^*_{G_0\times S^1}(P)\simeq H^*_{S^1}(M).$
\begin{defn}
  The $(G_0,S^1)$--equivariant vertical volume element
  $$
  \gamma_{G_0,S^1}\in\Omega(P)\otimes S(\g_0)\otimes\C[[m]],
  $$
  is defined by
  $$
  \gamma_{G_0,S^1}=\ex{(\Omega-\calI)\otimes\lambda}
  \calB_\eta\left(\ex{\theta\otimes\eta}\right),
  $$
  where $\theta$ is a $S^1$-equivariant $G$-connection on $P$,
  $\Omega_{S^1}=\Omega-\calI$ is the $S^1$-equivariant curvature, and
  $\calI=m\theta(X)$, where $X$ is the vector field corresponding to a
  generator $m$ of the Lie algebra $\uu(1)$ of $S^1$.
\end{defn}
\begin{prop}
  The following hold:
  \begin{enumerate}
  \item The functional $\Omega_{G_0,S^1}(P)\to\C[[m]]$ defined by
    $$
    \alpha\mapsto\int_P\ip{\alpha\wedge\gamma_{G_0,S^1}},
    $$
    descends to $H^*_{G_0,S^1}(P)$.
  \item If $\alpha^P\in H^*_{G_0,S^1}(P)$ and $\alpha^M\in H^*_{S^1}(M)$
    satisfy $\alpha^M=\text{ Hor}\circ CW_H(\alpha^P),$ then
    $$
    \int_P\ip{\alpha^P\wedge\gamma_{G_0,S^1}}=\int_M\alpha^M.
    $$
  \end{enumerate}
\end{prop}
A detailed proof of the proposition above is given in~\cite[p. 38-39]{radu}.
\begin{rem}
  \label{mreal}
  We can regard $m\in\uu(1)^*$ as a real number, so we can define $d_{G_0,m}$
  on $\Omega_{G_0}(P)^{S^1}$ by:
  $$
  d_{G_0,m}(\omega)=d\omega-m\iota_X\omega.
  $$
  Then $d_{G_0,m}^2=0,$ and we denote the corresponding cohomology group
  by $H^*_{G_0,m}(P)$. In this context, we define the corresponding vertical
  volume element by:
  $$
  \gamma_{G_0,m}\in\Omega(P)\otimes S(\g_0).
  $$
\end{rem}

\begin{prop}~\cite[p. 39]{radu}
  The functional $\Omega_{G_0\times S^1}(P)\to\C[m]$, defined by
  $$
  \alpha\mapsto\int_P\ip{\alpha\wedge\gamma_{G_0,m}},
  $$
  induces a functional integral
  $\displaystyle\int:H^*_{G_0,m}(P)\to\C[m]$.
\end{prop}


\subsection{Intersection numbers and localization on the moduli space}
\label{mqintnumbers}

In this section we describe the fundamental formulas which express the
relationship between the MQ formalism applied in a TQFT setup and topological
invariants of manifolds. We note here that the TQFT frameworks make use
of infinite dimensional configuration spaces, therefore one has to formally
extend the MQ construction to infinite dimensions. 

Let $N\stackrel{i}{\inc}M$ be a submanifold of a given finite-dimensional
manifold $M$, and $\alpha\in H^*(M)$. Then
$$
\int_N i^*\alpha=\int_M\alpha\wedge\eta,
$$
where $\eta$ is the Poincar\'e dual of the inclusion of $N$ in $M$.  If
$N_1,\dots,N_m$ are $m$ submanifolds of $M$ which intersect transversally,
and the sum of the codimensions of $N_i$ in $M$ equals the dimension of $M$,
then:
$$
\int_M\eta_1\wedge\dots\wedge\eta_m=\#(N_1\cap\dots\cap N_m).
$$
For a generic section $s$, the {\em localization principle} on the moduli
space $\calZ(s)$ states the following~\cite{cordes}:
\begin{equation}
  \label{locmodulispace}
  \int_M s^*(Th_t(E))\wedge\calO=\int_{\calZ(s)}i^*\calO,
\end{equation}
where $\calO$ is a product of observables (differential forms) of a given
TQFT.  This follows from a Poincar\'e duality argument.
\begin{prop}[\cite{cordes}]
  The following facts hold:
  \begin{enumerate}
  \item $Th_t(E)$ does not depend on $t$.
  \item If $t\to 0$ the support of $s^*Th_t$ becomes concentrated on
    $\calZ(s)$. 
  \item The stationary phase approximation of the integral
    $$
    \int_M s^*(Th_t(E))\wedge\calO
    $$
    is given by an integral over $\calZ(s)$ and a Gaussian integral in the
    normal directions. By {\em 1.}, the Gaussian approximation is exact.
  \end{enumerate}
\end{prop}
From the equivariant setup of the MQ formalism, the correlation functions of
the theory, for observables $\calO_i$, $1\le i\le k$, are described by:
\begin{eqnarray*}
  \ip{\calO_1\dots\calO_k}&=&(2\pi)^{-\dim\g}(2\pi t)^{-m}
  \int_P\hat{\calO}_1\wedge\dots\wedge\hat{\calO}_k\wedge
  \ex{-\frac{t}{2}\norm{s}^2}\calB_\eta\left(\ex{(C^*,\eta)}\right)\cdot\\
  &&\cdot\int_{\lambda,\phi\in\g}
  \ex{i\ip{\phi,C^*C\lambda}-idC^*\otimes\lambda}
  \calB_\chi\left(\ex{-itds\chi+\frac{t}{2}\chi\phi\chi)}\right)\\
  &=&(2\pi)^{-\dim\g}\int_{P,\g}\hat{\calO}_1\wedge\dots\wedge\hat{\calO}_k\wedge
  \Gamma_{G,t}\wedge U_t(s),
\end{eqnarray*}
where $\hat{\calO}_i$ are the images of the regular cohomology classes $\calO_i$ on
$M=P/G$ in the equivariant cohomology of $P$, assuming that $G$ acts freely.

The main obstruction to the MQ formalism in infinite dimensions is that
$e(E)$ is not well defined. Using the Mathai-Quillen formalism, one defines
an analogous Euler class for $E$.  The outcome of the construction is called
a {\em regularized Euler number} for the bundle $E$~\cite{aj}.
Unfortunately, it depends explicitly on the section chosen for the
construction, so it is important to make a good choice. In this context,
topological invariants are invariants of numbers like $\chi_s(E)$ under
deformations of certain data entering into their calculation. It is in this
sense that the Donaldson invariants of four-manifolds, which arise as
correlation functions of the TQFT considered in~\cite{ewdon}, are
topological, as they are independent of the metric which enters into the
definition of the instanton moduli space. As is well known, they have the
remarkable property of distinguishing differentiable structures on
four-manifolds.


\section{The Mathai--Quillen Formalism in TQFT} 
\label{mq_tqft}

The Mathai--Quillen (MQ) formalism can be used to give formal geometric
interpretations of the partition functions of certain TQFT (see
e.g.~\cite{radu},~\cite{labastida_marino3},~\cite{labastida_marino4},~\cite{marino},~\cite{wu}).
The general idea is that the partition function of TQFT is a path-integral
representation of the equivariant Euler number of a certain
infinite-dimensional bundle, and the path integrals describing the
correlation functions can be understood as representing intersection numbers
of equivariant cohomology classes.  Below we summarize the MQ formalism for
Donaldson--Witten (DW) and Seiberg--Witten (SW) TQFT.


\subsection{Donaldson--Witten Theory via MQ}
\label{dwmq}

The framework for Donaldson-Witten theory is the following. Let $X^4$ be a
simply-connected oriented 4-manifold with a Riemannian metric $g$.  Let $E$
be a rank 2 hermitian vector bundle over $X$ with structure group $SU(2)$ and
second Chern class $c_2(E)=k$. We denote by $\calA_E$ the affine space of
unitary connections on $E$, and by $\calG_E$ the group of unitary gauge
transformations of $E$ with Lie algebra $\g_E=su(E)$, the bundle of traceless
skew-Hermitian endomorphisms of $E$.

The setup for Mathai--Quillen formalism applied to Donaldson--Witten theory
is the following. The notation below was explained in $\S$\ref{mq}.
\begin{itemize}
\item[a)] $\calP=\calA_E$.
\item[b)] $\calG=\calG_E$.
\item[c)] $\calV=\Omega^2_+(\g_E)$.
\item[d)] $\calE=\calA_E\times_{\calG_E}\Omega^2_+(\g_E)$, an infinite
  dimensional bundle over $\calP/\calG$.
\item[e)] $s:\calP\to\calV,\quad s(A)=F_A^+$, which descends to a section
  $s:\calP/\calG\to\calE$.
\end{itemize}
The Mathai-Quillen formalism expresses the $\calG$-equivariant Euler number
$\chi_{\calG}(\calE,s)$ in terms of the section $s$. The formal expression
for the equivariant Euler number associated to DW theory is given by:
\begin{eqnarray}
  \label{dwpf}
  \chi_{\calG_E}(\calE,s)&=&\text{ const}\cdot\int_{\calA_E}
  \ex{-\frac{t}{2}\norm{F_A^+}^2}
  \calB_\eta\left(\ex{-\ip{D_S^*\psi,\eta}}\right)\cdot\nonumber\\
  &\cdot&\int\int_{\lambda,\phi\in\g}\ex{i\ip{\phi,D_A^* D_A\lambda}}
  \cdot\ex{-i[\psi,\psi]_0\lambda}\cdot
  \calB_\chi\left(\ex{-it\ip{(D_A\psi)^+,\chi}+\frac{t}{2}\chi[\phi,\chi]}
  \right)\nonumber\\
  &=&\text{ const}\cdot\int\calD A\calD\psi\calD\eta\calD\lambda\calD\phi
  \calD\chi\cdot\ex{-\int_X\Tr{\calL_{DW}}},
\end{eqnarray}
where
\begin{equation*}
  \calL_{DW}=\frac{t}{2}(F_A^+)^2 + \psi\wedge *D_A\eta
  - i *\psi D_A^* D_A\lambda + i *[\psi,\psi]_0\lambda
  +it D_A\psi\wedge\chi - \frac{t}{2}\chi\wedge[\phi,\chi]
\end{equation*}
is the Lagrangian of the theory. For details see~\cite[Sec 3.1]{radu}. The
equivariant Euler number~\eqref{dwpf} is the equivariant partition function
$Z_k(m)$ of the theory.  One can show that the constant in the above formula
is
$$
\text{const }=(2\pi)^{-\dim{\Omega^0(su(E))/2}}\cdot
(2\pi)^{-\dim{\Omega^2_+(su(E))/2}}.
$$
It also follows from the MQ formalism that~\eqref{dwpf} can be written as
\begin{equation}
  \label{dwmqpf}
  Z_k(m)=\int_{\calA_E/\calG_E} e_{\calG_E}(\calE,s)\wedge\Gamma_{\calG_E},
\end{equation}
where $\Gamma_{\calG_E}$ is a vertical volume element along the fibers, and
$e_{\calG_E}(\calE,s)$ is the equivariant Euler class.

\begin{rem}
  The MQ formalism for the DW theory is constructed for $b_2^+(X)>1$. Such a
  condition on $X$ assures that the localization on the moduli space formula
  (see $\S$\ref{mqintnumbers}) holds. For $b_2^+(X)=1$ there are choices
  involved in the construction of the DW moduli spaces, which are reflected
  in the path integrals over the configuration spaces $\calA_E/\calG_E$.  The
  MQ formalism will depend on these choices (e.g. metric chambers) and
  correction terms need to be added, but one still hopes to obtain similar
  formulas for the DW partition function.  We have not yet worked out the
  details for this case.
\end{rem}
We enumerate below some important facts from Donaldson theory. 
\begin{prop}
  \label{propdimD}
  If the Betti numbers satisfy $b_2>0$ and $b_1=0$, then for a generic metric
  on $X$, there are no reducible anti-self-dual (ASD) connections in
  $\calM_k$, the moduli space of instantons on $E$, which is a finite
  dimensional manifold of dimension
  $$
  \dim{\calM_k}=8k-3(1+b_2^+)=8k-\frac{3}{2}(\chi+\sigma),
  $$
  where $\chi$ and $\sigma$ are the Euler characteristic and the
  signature of $X$, respectively.
\end{prop}

\begin{rem}
  The reducible connections satisfy the following:
  \begin{enumerate}
  \item $A\in\calA_E$ is reducible \iff preserves a splitting
    $$
    E=\lambda_1\oplus\lambda_2.
    $$
    This is equivalent to either $\hbox{Stab}(A)/\Z_2\simeq U(1)\simeq
    S^1$, where $\hbox{Stab}(A)$ is the stabilizer of $A$ and $\Z_2$ is the
    center of $\calG_E$, or $\ker d_A$ is 1-dimensional.  We denote by
    $\calA^*_E$ the space of irreducible connections. Then $\calG_E/\Z_2$
    acts freely on $\calA^*_E$.
  \item For $E$ an $SU(2)$-bundle, $A$ is reducible \iff $A$ preserves a
    splitting $E=l\oplus l\inv$, so $\g_E=\R\oplus l^2$. Moreover we get an
    action of $\hbox{Stab}(A)$ on $\calA_E$ by the square of the standard
    action.  Note that
    $$
    E=l\oplus l\inv\Rightarrow c_2(E)=-c_1(l)^2.
    $$
  \item There is an elliptic complex
    $$
    0\to\Omega^0(\ad{\calE})\stackrel{d_A}{\to}\Omega^1(\ad{\calE})
    \stackrel{P_{\pm}d_A}{\to}\Omega^2_{\pm}(\ad{\calE})\to 0,
    $$
    so $\ker{d_A}=H^0_A$ of the above complex.  The elliptic complex is
    $\text{Stab}(A)$-equivariant, so $\text{Stab}(A)$ acts on $H^*_A$, where
    $H_A^*$ is the cohomology of this complex.  Moreover, $\text{Stab}(A)$
    acts trivially on $H^0_A$.
  \item The set of irreducible ASD connections modulo the gauge group
    $\calG_E$ is in 1-1 correspondence with the set
    $$
    \{(\pm c,\pm c)\,|\, 0\ne c\in H^2(X;\Z), c^2=-c_2(E)\},
    $$
    with holonomy group $S^1$ (see~\cite[Prop 4.2.15]{dk}).
  \end{enumerate}
\end{rem}
The DW polynomial invariants are defined as follows.  Let $\Sigma\in
H_2(X,\Z)$, and define
$$
\hat{\mu}=\hat{\mu}^{2,0}(\Sigma)+\hat{\mu}^{0,1}(\Sigma)
\in\Omega^2(\calA_E)\oplus\left(\Omega^0(\calA_E)\otimes S^1(\g_E^*)
\right),
$$
where
$$
\hat{\mu}^{2,0}(\Sigma)(\psi_1,\psi_2)=\frac{1}{8\pi^2}
\int_\Sigma\Tr(\psi_1\wedge\psi_2),\quad\text{for }\psi_1,\psi_2\in T_A\calA_E,
$$
$$
\hat{\mu}^{0,1}(\Sigma)(A,\phi)=\frac{1}{4\pi^2}
\int_\Sigma\Tr(\phi F_A),\quad\text{for }\phi\in\g_E, A\in\calA_E.
$$
For $\nu\in H_4(X,\Z)$ define:
$$
\hat{\mu}(\nu)\in S^2\left(\g_E^*\right),\quad
\hat{\mu}(\nu)=\frac{1}{8\pi^2}\int_\nu\Tr(\phi^2),
\quad\text{for }\phi\in\g_E.
$$

One can show that $\hat{\mu}(\Sigma), \hat{\mu}(\nu)$ are $\calG_E$-invariant,
and hence define equivariant differential forms in $\Omega^2_\calG(\calA_E)$,
and $\Omega^4_\calG(\calA_E)$, respectively.  Moreover,
$\hat{\mu}(\Sigma), \hat{\mu}(\nu)$ are equivariantly closed, and thus define
classes in $H^*_\calG(\calA)$.

\begin{rem}
  When restricted to $\calA^*_E$, the space of irreducible connections, 
  $\hat{\mu}(\Sigma)$ and $\hat{\mu}(\nu)$ determine classes in
  $H_\calG(\calA^*_E)$. The action of $\calG_E$ on $\calA^*_E$ is free,
  so the Chern--Weil map
  $$
  CW:H_\calG(\calA^*_E)\to H(\calA^*_E/\calG_E)
  $$
  applied to $\hat{\mu}(\Sigma)$ and $\hat{\mu}(\nu)$ yields
  $$
  \mu(\Sigma)\in H^2(\calA^*_E/\calG_E),\qquad\mu(\nu)\in H^4(\calA^*_E/\calG_E).
  $$
  Note that $\mu(\nu)=c_2(\calU)\!\mid_{\calA_E/\calG_E}$, and
  $\mu(\Sigma)$ is the slant product of $[\Sigma]$ and $c_2(\calU)$, where
  $\calU$ is the universal bundle over $X\times\calA_E/\calG_E$.
\end{rem}

The $k$-th Donaldson polynominal,
$\D_k:\R[\Sigma_1,\dots,\Sigma_{b_2},\nu]_{(k)}\to\R$, where $\Sigma_i$ have
degree 2, $\nu$ has degree 4, and $(k)$ denotes polynomials of total degree
$$
d(k)=8k-3(1+b_2^+),
$$
is defined by:
\begin{equation}
  \label{donaldson_k}
  \D_k(\Sigma_1^{\alpha_1},\dots,\Sigma_{b_2}^{\alpha_{b_2}},\nu^\beta)
  =\int_{\calM_k}\mu(\Sigma_1)^{\alpha_1}\dots
  \mu(\Sigma_{b_2})^{\alpha_{b_2}}\mu(\nu)^\beta.
\end{equation}
$\D_k$ is a polynomial of degree $d(k)$ on $H^2\oplus H^0$.  To be rigorous,
one has to consider the Uhlenbeck compactification of the moduli spaces
$\calM_k$, choose suitable representatives for $\mu$'s which extend to the
compactification, prove the independence of the choices, etc.

The total Donaldson polynomial is defined by
$$
\D_X(\Sigma_1^{\alpha_1},\dots,\Sigma_{b_2}^{\alpha_{b_2}},\nu^\beta)
=\sum_{k\in\Z}\int_{\calM_k}\mu(\Sigma_1)^{\alpha_1}\dots
\mu(\Sigma_{b_2})^{\alpha_{b_2}}\mu(\nu)^\beta.
$$
\begin{rem}
  \label{thetwofactor}
  Strictly speaking, the instanton number $k$ must be positive.  Formally,
  allowing $k$ to have also negative values, comes down to a factor of 2 in
  the definition of the Donaldson invariant.  The difference comes from a
  normalization argument (see e.g.~\cite[(2.18)]{ewdon2}). From this point
  on, we omit the factor of 2, and we consider $k\in\Z$. In
  Section~\ref{conjecture} we will add the factor of 2, in order to agree to
  the usual topological conventions, and we will refer to this remark.
\end{rem}
\noindent If we introduce formal variables $q_1,\dots,q_{b_2},\lambda$, we can write
$$
\D_X((q_1\Sigma_1)^{\alpha_1},\dots,(q_{b_2}\Sigma_{b_2})^{\alpha_{b_2}},
(\lambda\nu)^\beta)=\sum_{k\in\Z}q_1^{\alpha_1}\dots q_{b_2}^{\alpha_{b_2}}
\lambda^\beta\cdot\int_{\calM_k}\mu(\Sigma_1)^{\alpha_1}\dots
\mu(\Sigma_{b_2})^{\alpha_{b_2}}\mu(\nu)^\beta.
$$
Using generating function notation~\cite{ewdon}, we set
$$
\D_X\left(e^{\sum q_a\Sigma_a+\lambda\nu}\right)=
\sum \frac{\D_X((q_1\Sigma_1)^{\alpha_1},\dots,(q_{b_2}
  \Sigma_{b_2})^{\alpha_{b_2}},(\lambda\nu)^\beta)}{\alpha_1!\cdots
  \alpha_{b_2}!\cdot\beta!}.
$$
For simplicity, we will work from now on with observables $\calO$
containing only one $\Sigma$. Then the generating function can be written in
terms of a sum over the dimension $d=d(k)$ of the moduli spaces $\calM_k$.
By~\eqref{donaldson_k}, we obtain:
\begin{equation}
  \label{dwgeneratingfunction}
  \D_X\left(e^{\Sigma+\lambda\nu}\right)=
  \sum_{d}\sum_{a+2b=2d}\frac{(2\lambda)^b}{a!b!}
  \int_{\calM_k}\mu(\Sigma)^{a}\mu(\nu)^b.
\end{equation}
If $\calO$ is product of $\mu$ classes, the MQ formalism provides the
following expression for the total equivariant Donaldson invariant:
$$
\D_X(\calO)(m)=\text{ const }\cdot\sum_{k\in\Z}\int_{\calA_k}\int_{\g_k}
\Gamma_{\calG_k}\wedge U_{\calG_k,t}(s)\wedge\calO,
$$
where for simplicity of notation, we denote $\calA_{E_k}$ by $\calA_k$,
$\calG_{E_k}$ by $\calG_k$, etc.  More concretely:
$$
\D_X(\Sigma^{\alpha},\nu^\beta)(m) =\text{ const }
\cdot\sum_{k\in\Z}\int_{\calA_k,\g_k}
\calD(\text{fields})\wedge\hat{\mu}(\Sigma)^{\alpha}\wedge\hat{\mu}(\nu)^\beta
\cdot\ex{-\int_X\Tr{\calL_{DW}}}.
$$
Summarizing, one obtains the formal equality of generating series:
\begin{eqnarray*}
  \D_X(e^{q\Sigma+p\nu})(m)&=&\text{ const }
  \cdot\sum_{k\in\Z}\int_{\calA_k,\g_k}\calD(\text{fields})\cdot\\
  &&\cdot\ex{-\int_X\Tr{\calL_DW}+\frac{p}{8\pi^2}\int_X\Tr{\phi^2}
    +\frac{q}{8\pi^2}\int_\Sigma\Tr(\psi\wedge\psi+2\phi F_A)}.
\end{eqnarray*}
The Donaldson simple type condition is defined as follows:
\begin{defn}
  For $b_2^+\ge 2$ and odd, $X$ is said to be of {\em D-simple type}
  if for every $z\in A(X)=S^*(H_0(X)\oplus H_2(X))$,
  $$
  \D_X(u^2z)=4\D_X(z).
  $$
\end{defn}
\begin{rem}
  The condition of D-simple type above can be reformulated in the form that
  $\D_X$ annihilates the ideal in $A(X)$ generated by $u^2-4$. Equivalently,
  $$
  \D_X(e^{\lambda u}z)=e^{2\lambda}
  \D_X\left(\left(1+\frac{u}{2}\right)z\right)+
  e^{-2\lambda}
  \D_X\left(\left(1-\frac{u}{2}\right)z\right),
  $$
  for $z\in A(X)$ and $\lambda\in\Z$.
\end{rem}
Assuming the D-simple type condition, the generating series can defined as
in~\cite{km}:
$$
\D_X(\exp(\calO))=\sum_{d'}\frac{q_{d'}}{(d')!},
$$
where the Donaldson polynomials of degree $d'=d'(k)$ are defined by
$$
q_{d'-2}=\ip{\mu(\Sigma)^{d'-2}\nu,[\calM_{d'}]}.
$$
Here $d'$ is considered as
$$
d'\equiv\frac{1}{4}(\chi+\sigma)\mod 2,
$$
rather than $\mod 4$. Equivalently, one can consider $k\in\frac{1}{2}\Z$.

\subsection{Seiberg-Witten Theory via MQ}
\label{swmq}

The framework is the following. Let $X^4$ be a closed oriented manifold with
Riemannian metric $g$, equipped with a \spinc-structure $c$. Let $W^\pm$ be
the corresponding rank 2 hermitian vector bundles (the spin bundles), and $L$
the hermitian line bundle over $X$ with $c_1(L)=c$, and $\det(W^+)=L$.
Let $\rho:\Lambda^2_+(X)\to su(W^+)$ be a bundle isomorphism, and
$\eta\in\Omega^2_+(X)$ a perturbation. We denote by $\calA_L$ the space of
unitary connections on $L$.

The Seiberg-Witten (SW) equations for a connection $A\in\calA_L$ and a spinor
$\psi\in\Gamma(W^+)$ are:
\begin{eqnarray*}
  \Dirac_A\psi&=&0\\
  F_A^+ + i(\psi\otimes\bar{\psi})_{oo}&=&0,
\end{eqnarray*}
where $\Dirac_A$ is the Dirac operator associated to the \spinc-structure and
the connection $A$, $i(\psi\otimes\bar{\psi})_{oo}\in\Gamma(su(W^+))$ is the
traceless part of $\psi\otimes\bar{\psi}$, and $su(W^+)$ and $\Lambda^2_+(X)$
are isomorphic as Clifford algebras, via $\rho$.  One can consider a
perturbation of the second equation with the form $\eta$.

The MQ formalism for the SW theory has the following setup:
\begin{itemize}
\item[a)] $\calP=\calA_L\times\Gamma(W^+)$.
\item[b)] $\calV=\Omega^2_+\oplus\Gamma(W^-)=:\calV_1\oplus\calV_2$.
\item[c)] $\calG_L=\hbox{Aut}(L)=\Map(X,U(1))$.
\item[d)] $\calE=\calP\times_{\calG}\calV$, and
  $\calE_1=\calP\times_{\calG}\calV_1$, 
  $\calE_2=\calP\times_{\calG}\calV_2$. 
\item[e)] $s:\calP\to\calV,\quad
  s(A,\psi)=(F_A^+ + i(\psi\otimes\bar{\psi}),\Dirac_A\psi)=:(s_1,s_2)$. 
\end{itemize}
Here $\calG_L$ acts by the usual multiplication by $\ex{i\theta}$ in the
fibers of $W^+$, and $\calG_L$ acts on $\calA_L$ through the gauge
transformation $\ex{2i\theta}$, i.e.
$$
g\cdot(A,\psi)=(A-2g\inv dg, g\psi).
$$
We enumerate below some important aspects of SW theory. The moduli space
$\calM_c$ is the space of solutions of the SW equations modulo the gauge
group $\calG_L$, and it is proven to be a finite dimensional, compact,
oriented, smooth manifold for a generic metric and regular perturbation, of
dimension
\begin{equation}
  \label{propdimSW}
  \dim\calM_c=\frac{c^2-2\chi-3\sigma}{4}
\end{equation}
The rolled-up SW elliptic complex is
$$
0\to\Lambda^1\oplus\Gamma(X,W^+\otimes L)\stackrel{\Dirac_A+d^+ +d^*}{\to}
\Lambda^0\oplus\Lambda^{2+}\oplus\Gamma(X,W^-\otimes L)\to 0.
$$
The SW equations are gauge invariant, and so $s$ is $\calG_L$-equivariant.
The action of $\calG_L$ is not free. If $\psi\ne 0$ the stabilizer of
$(A,\psi)$ is trivial, and if $\psi=0$ the stabilizer of $(A,0)$ is $S^1$.
For $x_0\in X$ we set
$$
\calG_L^0=\{g\in\calG_L\mid g\cdot x_0=1\}.
$$
Then $\calG_L=\calG_L^0\times S^1$ and $\calG_L^0$ acts freely on
$\calA_L\times\Gamma(W^+)$.

The SW invariant $SW(c)$ for $\dim\calM_c=0$ is defined to be the sum of $\pm
1$ over the points in the moduli space, according to their orientation.  For
$\dim\calM_c>0$, the SW invariant is defined as follows.  Consider the
principal $S^1$ bundle $\calM_{\calG_L^0}\to\calM_c$ given by
$$
\calM_{\calG_L^0}=s\inv(0)/\calG_L^0\subset\calP/\calG_L^0,
$$
for a generic metric, and perturbation. Equivalently, we can consider the
associated universal line bundle
$$
\calL\to X\times P
$$
restricted to ${x_0}\times P$. Then~\cite{salomon}
$$
e(\calM_{\calG_L^0}\to\calM_c)=c_1(\calL\big|_{x_0}).
$$
If $\dim\calM_c=2s(c)$, the SW invariant is defined to be
$$
SW(c)=\int_{\calM_c} c_1(\calL\big|_{x_0})^s
$$

The computation of the quantities in the MQ construction for
the unperturbed equations can found in~\cite[Sec. 3.2]{radu}.
One can write:
\begin{eqnarray*}
  U_t(s)&=&\text{ const }\cdot\int\calD\chi\calD T
  \ex{t\left(\int_X-\frac{1}{2}(F^+_A)^2-
      F^+_A\wedge i(\psi\otimes\bar{\psi})_0
      -\frac{1}{2}\norm{i(\psi\otimes\bar{\psi})_0}^2
      -\frac{1}{2}*\norm{\Dirac_A\psi}^2\right)}\cdot\\
  &&\ex{t\left(\int_X -i\left(d^+\alpha+i(\psi\otimes\bar{\sigma}
        +\sigma\otimes\bar{\psi})_0\right)\chi
      -i\ip{\not\!D_A\sigma+\frac{1}{2}\cl(\alpha)\psi,T}
      +\frac{1}{2}\ip{T,\phi T}\right)}
\end{eqnarray*}
for all $t>0$. The constant is given by
$$
\text{const }=(2\pi)^{-\dim{\Omega^0}/2}
(2\pi t)^{-\dim(\Omega^2_+\otimes\Gamma(W^-))/2}.
$$
Finally, the equivariant Euler number (the SW partition function) of the
theory is given by the following formula:
$$
Z_c(m)=\text{ const }\cdot\int_{\calA_L\times\Gamma(W^+)\times\Map(X,u(1))}
\calD A\calD\psi\calD\phi\,\,\Gamma_{\calG_L}\wedge U_t(s),
$$
where $\Gamma_{\calG_L}$ is a vertical volume element along the fibers.  By
integrating over the fiber, one gets the equivalent expression
$$
SW(c,m)=\int_{\calP/\calG} e_{\calG_L}(\calE,s)\wedge\Gamma_{\calG_L}.
$$
\begin{rem}
  For $c^2=2\chi+3\sigma$, the partition function gives a formal expression
  for the equivariant SW invariant $SW(c,m)$.  For $\dim\calM_c=2s(c)>0$, we
  can define the equivariant SW invariants in terms of the equivariant first
  Chern class of the universal bundle $\calL\big|_{x_0}$:
  $$
  SW(c,m)=\int_{\calP/\calG} e_{\calG}(\calE,s)\wedge\Gamma_{\calG_L}
  \wedge c_{1, \calG_L}(\calL\big|_{x_0})^s.
  $$
\end{rem}


\section{Equivariant Localization} 
\label{equivloc}

We recall below the abelian localization theorem, in the case of an $S^1$
action on a given manifold, and then we formally apply an infinite version of
it to the DW and SW theories.  We note here that there are more general
versions of localization formulas for non-abelian group actions (see
e.g.~\cite{bgv},\cite{guillemin}).

The setup is the following.  Let $M$ be a compact oriented manifold equipped
with an $S^1$ action and a Riemannian metric $g$ invariant under the action.
Let $\alpha\in\Omega_{S^1}(M)$ be an equivariantly closed form, and $m$ a
generator of $\uu(1)^*$ which induces the vector field $X$ on $M$.  Denote by
$M_0$ the fixed point set of the $S^1$ action (the zero set of $X$).  The
abelian localization theorem~\cite[Theorem 7.13]{bgv} asserts:
\begin{thm}
  $M_0$ is a submanifold of $M$ whose normal bundle $\nu_{M_0}$ is
  orientable and even dimensional. Moreover,
  \begin{equation}
    \label{abloc}
    \int_M\alpha=\int_{M_0}\frac{\alpha\big|_{M_0}}{e_{S^1}(\nu_{M_0})}.
  \end{equation}
  where $\alpha\big|_{M_0}=i^*\alpha$ for the inclusion map $i:M_0\inc M$.
\end{thm}
On the left hand side of~\eqref{abloc}, 
$\alpha\in\C[m]\otimes\Omega(M)^{S^1}$, so 
$$
\int_M\alpha\in\C[m].
$$
On the right hand side, we have:
$$
e_{S^1}(\nu_{M_0})=m^{\frac{k}{2}}e^0 + m^{\frac{k}{2}-1}e^2 +\cdots+ e^k
\in H_{S^1}^{k/2}(M_0),
$$
where $k=\rk(\nu_{M_0})$ and $e^i$ are elements of $H^i(M_0)$.
Then
$$
e_{S^1}(\nu_{M_0})\inv=\frac{1}{m^{\frac{k}{2}}e^0}\left(
  1+\sum_{k\ge 1}\left(\frac{1}{e^0}(m\inv e^2+\cdots+m^{-\frac{k}{2}}e^k)
  \right)^k\right)
$$
is a well defined homogenous element of degree $-k/2$ in
$$
\Omega_{S^1}(\nu_{M_0})_m=\C[m,m\inv]\otimes\Omega(M_0)^{S^1}.
$$
Here $\deg{m}=2$ and $\alpha\mid_{M_0}\cdot e_{S^1}(\nu_{M_0})\inv$ is a
homogenous element of degree $(\deg{\alpha}-k)/2.$ Note that
$\displaystyle\int_{M_0}$ picks up those terms whose {\em usual} degree as a
form is $\dim{M_0}$. The result is a multiple of $m^{(\deg{\alpha}-k)/2}.$

It is important to remark here that we can regard $m$ as a real number (see
Remark~\ref{mreal}). In order to keep track of the degree 2 of $m$, we must
think of $m$ as actually being a square of a real number.


\subsection{Abelian Localization of Donaldson--Witten theory}
\label{dwabloc}

The equivariant DW partition function $Z_k(m)$ for $c_2(E)=k$ can be written
as an integral over the configuration space:
\begin{equation}
  \label{dwpf1}
  Z_k(m)=\int_{{\calA}_E/{\calG}_E} 
  e_{\calG_E}(\calE,s)\wedge\Gamma_{\calG_E}\in\C[m].
\end{equation}
Consider the based gauge group $\calG_0=\{g\in\calG_E | gx=\id\}$, for fixed
$x\in X$. We have the following exact sequence:
$$
1\to\calG_0\to\calG_E\to SU(2)\to 1.
$$
Using the action of a maximal torus $S^1$ of $SU(2)$ we can
rewrite~\eqref{dwpf1} in a form more suitable for abelian localization.  At
this moment we assume that~\eqref{abloc} holds in infinite dimensions, as
well.  We get:
\begin{equation}
  \label{dwablocthm}
  Z_k(m)=\int_{{\calA}_E/{\calG}_E} e_{S^1}(\calE,s)\wedge\Gamma_{S^1}
  =\int_{M_0/\calG_E} e_{S^1}(\calE,s)\big|_{M_0}
  \wedge\Gamma_{S^1}\big|_{M_0}
  \wedge e_{S^1}(\nu_{M_0})\inv,
\end{equation}
where we first apply~\eqref{abloc} and then mod out by $\calG_E$.  Note that
$S^1$ acts on $\Omega^1(\g_E)$ by the square of the standard action, and
$\calG_E/\Z_2$ acts freely on $\calA^*_E$.  The fixed point set for the
$S^1$ action is given by the {\em reducible} connections. For $E=l\oplus
l\inv$ we have
\begin{equation}
  \label{redconn}
  \g_E=\R\oplus l^2\quad\text{and}\quad d_{A_E}=d\oplus d_{A_{l^2}}.
\end{equation}
The original DW elliptic complex is 
$$
0\to\Omega^0(\g_E)\stackrel{d_{A_E}}{\to}
\Omega^1(\g_E)\stackrel{d^+_{A_E}}{\to}
\Omega^{2+}(\g_E)\to 0,
$$
and by~\eqref{redconn} the elliptic
complex splits in two elliptic complexes (see~\cite[p. 70]{fu}):
\[
\begin{diagram}
  \node{0}\arrow{e}
  \node{\Omega^0(l^2)}\arrow{s,l}{\oplus}\arrow{e,t}{d_A}
  \node{\Omega^1(l^2)}\arrow{s,l}{\oplus}\arrow{e,t}{d_A^+}
  \node{\Omega^{2+}(l^2)}\arrow{s,l}{\oplus}\arrow{e}
  \node{0}\\
  \node{0}\arrow{e}
  \node{\Omega^0}\arrow{e,t}{d}
  \node{\Omega^1}\arrow{e,t}{d^+}
  \node{\Omega^{2+}}\arrow{e}
  \node{0}
\end{diagram}
\]
Let $\calA_l$ be the space of connections on $l$.  Then the fixed point set
of the $S^1$-action, is
$$
M_0=\bigcup_{E=l\oplus l\inv}\,\calA_{l^2} =\bigcup_{\stackrel{y\in
    H^2(X;\Z)}{y^2=-k}} \left(\bigcup_{c_1(l)=2y}\,\calA_{l}\right),
$$
where we use~\eqref{redconn} to identify a reducible connection on $\g_E$
with a connection on $l^2$, and hence on $l$.  The last equality uses the
fact that $X$ is simply connected, and hence $H^2(X,\Z)$ has no 2-torsion.
Now we define
$$
\calR_x=\bigcup_{c_1(l)=x}\,\calA_l.
$$
The spaces $\calR_x$ are $\calG_E$-invariant. 
Moreover $\Gamma_{S^1}\big|_{M_0}=1$~\cite[p. 67]{radu}.
Then~\eqref{dwablocthm} becomes
\begin{eqnarray*}
  Z_k(m)&=&\int_{M_0/\calG_E}
  e_{S^1}(\calE,s)\wedge 
  e_{S^1}(\nu_{M_0\subset\calA_E/\calG_E})\inv\\
  &=&\sum_{\stackrel{x=2y}{y^2=-k}}
  \int_{\calR_x/\calG_E} e_{S^1}(\calE,s)\wedge
  e_{S^1}(\nu_{M_0\subset\calA_E/\calG_E})\inv.
\end{eqnarray*}
From this point on we will write the sum in the formula above as a sum over
$x$, with $x^2=-4k$.  The diffeomorphism
$$
i:\calA_{l_x}/\calG_{l_x}\approx\calR_x/\calG_E
$$
yields the following expression for the partition function:
\begin{equation}
  \label{dwabloc1}
  Z_k(m)=\sum_{x^2=-4k}
  \int_{\calA_{l_x}/\calG_{l_x}}
  i^*\left(e_{S^1}(\calE,s)\wedge
    e_{S^1}(\nu_{M_0})\inv\right).
\end{equation}
Let $\nu_{\calR_x}$ be the normal bundle of $\nu_{\calR_x}$ in $\calA_E$. At
a point $A\in\calA_E$, the fiber of $\nu_{\calR_x}$ is isomorphic to
$$
\{a\in\Omega^1(l^2)|d_A^*a=0\}.
$$
Here $\nu_{\calA_l}\subset\calA_E$, and because $\g_E\simeq\R\oplus
l^2$, we have $\nu_{\calA_l}\simeq\Omega^1(l^2).$ The structure group of $E$
and $\nu_{M_0}$ is $\calG_E$. When pulled back by $i^*$ to
$\calA_{l_x}/\calG_{l_x}$, their structure group reduces to $\calG_{l_x}$.
Then, for every $l=l_x$, we get:
\begin{eqnarray*}
  i^*\calE&=&\calA_l\times_{\calG_l}\Omega^2_+(\g_E)=
  \calA_l\times_{\calG_l}\left(\Omega^2_+\oplus
    (\Omega^2_+\otimes l^2)\right)\\ 
  i^*(\nu_{M_0})&=& \calA_l\times_{\calG_l}\nu_{\calA_{l_x}}=
  \calA_l\times_{\calG_l}\{a\in\Omega^1(l^2) | d_A^*a=0\}
  =\calA_l\times_{\calG_l}\Omega^1(l^2)/d_A\Omega^0(l^2).
\end{eqnarray*}
We decompose $i^*\calE$ as follows:
\begin{equation*}
  i^*\calE=\left(\calA_l\times_{\calG_l}\Omega^2_+\right)\oplus
  \left(\calA_l\times_{\calG_l}(\Omega^2_+\otimes l^2)\right)
  =:\calE_1\oplus\calE_2
\end{equation*}
For $\calG_l^0\subset\calG_l$ the based group, we have
$$
1\to\calG_l^0\to\calG_l\to S^1\to 1,
$$
and because $S^1$ acts trivially on $\calA_l/\calG_l$, we get the isomorphism
$$
\calA_l/\calG_l\simeq\calA_l/\calG_l^0.
$$
Moreover, $S^1$ acts on $\calE_1$ trivially, and $S^1$ acts on $i^*\nu$ with
weight 2. Then
$$
e_{S^1}(\calE_1,s)=e(\calE_1,s),
$$
so~\eqref{dwabloc1} becomes
$$
Z_k(m)=\sum_{x^2=-4k}\int_{\calA_{l_x}/\calG_{l_x}}
e(\calE_1,s)\wedge e_{S^1}(\calE_2,s)\wedge
e_{S^1}(i^*\nu_{M_0})\inv.
$$
More precisely,
\begin{eqnarray}
  \label{dwpf2}
  Z_k(m)&=&\sum_{x^2=-4k}\int_{\calA_{l_x}/\calG_{l_x}}
  e(\calA_l\times_{\calG_l}\Omega^2_+,s)\wedge 
  e_{S^1}(\calA_l\times_{\calG_l}(\Omega^2_+\otimes l^2),s)\nonumber\\
  &&\qquad\qquad\qquad\wedge\, 
  e_{S^1}(\calA_l\times_{\calG_l}\Omega^1(l^2)/d_A\Omega^0(l^2))\inv.
\end{eqnarray}
Since the kernel and cokernel of the operator
$$
d_A^* + d_A^+ :\ker{d_A^*}\to\Omega^{2+}(l^2)
$$
are isomorphic to the kernel and cokernel of the elliptic operator
$d_A^*+d_A^+$ defined on $\Omega^1(l^2)$, we can compute the quotient of
Euler classes in~\eqref{dwpf2} in terms of the total Chern class of (minus)
the index bundle of $d_A^*+d_A^+$. For a similar argument see~\cite[Prop
4.4]{radu}.  Assuming that this finite dimensional formula holds in infinite
dimensions, we obtain:
\begin{equation*}
  \frac{e_{S^1}(\calA_l\times_{\calG_l}(\Omega^2_+\otimes l^2),s)}
  {e_{S^1}(\calA_l\times_{\calG_l}\Omega^1(l^2)/d_A\Omega^0(l^2))}=
  \left(-\frac{4m}{2\pi}\right)^{-\Ind(d_A^++d_A^*)}
  c_{\text{tot}}\left(-\ind(d_A^++d_A^*)\right)\left(-\frac{2\pi}{4m}\right),
\end{equation*}
where $m$ is the generator of $\uu(1)^*$, $\Ind$ denotes the integer
index, and the factor of 4 comes from the $S^1$ action of $L^2$.  The
splitting of the elliptic complex into two elliptic complexes shows that
$$
\Ind(d_A^++d_A^*)=-\Ind(d^++d^*)+2d(k),
$$
where $2d(k)=\dim\calM_k$. When $2d(k)=0$, we can check that the (real)
indexes agree:
$$
\Ind(d_A^++d_A^*)=8k-(\chi+\sigma)=\frac{1}{2}(\chi+\sigma)=-\Ind(d^++d^*),
$$
since $0=2d(k)=8k-\frac{3}{2}(\chi+\sigma)$. 

\begin{rem}
  \label{dimaspect}
  It important to notice here the following formal aspect. The degrees of the
  forms inside the integrals over the infinite dimensional spaces
  $\calA_{l_x}/\calG_{l_x}$ are considered to sum to the (finite) dimension
  of the respective Donaldson moduli spaces. Therefore, one can consider the
  {\em formal dimension} of the integrals in question to equal the dimension
  of the corresponding moduli spaces. For this reason we only pick the degree
  zero component of the total Chern class above, denoted by $\1$.
\end{rem}
Using the remark above, we obtain:
\begin{equation}
  \label{dwabloc2}
  Z_k(m)=2^{-(\chi+\sigma)}\sum_{x^2=-4k}\int_{\calA_{l_x}/\calG_{l_x}}
  e(\calA_l\times_{\calG_l}\Omega^2_+,s)\wedge 
  \left(\frac{m}{2\pi}\right)^{-\frac{\chi+\sigma}{2}}\cdot\1
\end{equation}
where we notice that $-1$ is raised to an even power, so it equals 1.

The localization of (the equivariant version of) the observables of the
theory is given by:
$$
\mu(\Sigma)\big|_{M_0}=-\ip{c_1(l),\Sigma}\cdot h,
$$
where $h$ is the two-dimensional generator of the integer cohomology of
$BS^1$ (see~\cite[p. 187]{dk}). But we have seen that $H^*(BS^1)\cong
H^*_{S^1}(pt)=\C[m]$.  Therefore, $h$ can be identified with
$\displaystyle\left(\frac{m}{2\pi}\right)^2\cdot\1$. This gives
\begin{equation}
  \label{sigmarestr}
  \mu(\Sigma)\big|_{M_0}=-\ip{c_1(l),\Sigma}\cdot\left(\frac{m}{2\pi}\right)^2\cdot\1.
\end{equation}
Moreover, it is not hard to show that the restriction of $\mu(\nu)$ to the
fixed point set is given by (minus) the generator of the four-cohomology of
$BS^1$, thus
\begin{equation}
  \label{nurestr}
  \mu(\nu)\big|_{M_0}=-\left(\frac{m}{2\pi}\right)^4\cdot\1,
\end{equation}
Recall that the equivariant $k$-th Donaldson polynomial for
$\calO=\Sigma^a\wedge\nu^b$ is defined as follows:
\begin{equation}
  \D_k(\calO)(m)=\int_{{\calA}_E/{\calG}_E}
  e_{\calG_E}(\calE,s)\wedge\Gamma_{\calG_E}
  \wedge\mu(\Sigma)^a\wedge\nu^b,
\end{equation}
where $2d=2d(k)=a+2b$. Now we compute the localization formula for the
corellation functions of DW theory, applying~\eqref{sigmarestr}
and~\eqref{nurestr}, obtaining a formula for the equivariant Donaldson
polynomials in $\C[m]$.  The dependence of $m$ will denoted in parentheses,
as in the partition function case.
\begin{eqnarray}
  \label{dwobs1}
  &&\D_k(\calO)(m)=\sum_{x^2=-4k}\int_{\calA_{l_x}/\calG_{l_x}}
  e(\calA_{l_x}\times_{\calG_{l_x}}\Omega^2_+,F_A^+)\wedge
  \left(\frac{4m}{2\pi}\right)^{-\frac{\chi+\sigma}{2}-2d}\cdot\1\nonumber\\
  &&\qquad\qquad\qquad\qquad\wedge(-1)^{a}\ip{x,\Sigma}^{a}\cdot\left(\frac{m}{2\pi}\right)^{2a}\1
  \wedge(-1)^{2b}\left(\frac{m}{2\pi}\right)^{4b}\cdot\1\nonumber\\
  &=&\sum_{x^2=-4k}2^{-(\chi+\sigma)-4d}\ip{x,\Sigma}^{a}
  \int_{\calA_{l_x}/\calG_{l_x}} e(\calA_{l_x}\times_{\calG_{l_x}}\Omega^2_+,F_A^+)\wedge
  \left(\frac{m}{2\pi}\right)^{-\frac{\chi+\sigma}{2}+2d}\cdot\1\nonumber\\
  \,
\end{eqnarray}
\begin{rem}
  \label{dwrem}
  If we heuristically compare~\eqref{dwobs1} with~\eqref{dwabloc2}, we can
  formally write the $k$-th (equivariant) Donaldson polynomial
  $\D_k(\calO)(m)$ in terms of the partition function $Z_k(m)$, which
  represents the zero dimensional Donaldson invariant $q_0$. If we formally
  replace (modulo $2\pi$) $m$ by $m^{2d(k)}$, then we can write the
  equivariant Donaldson--Witten generating
  function~\eqref{dwgeneratingfunction} as follows:
  \begin{equation}
    \label{this}
    \D_X(e^{\Sigma+\lambda\nu})(m^{2d(k)})=\sum_k e^{\ip{x,\Sigma}+2\lambda} Z_k(m).
  \end{equation}
  This formula is similar to the one obtained by Witten
  in~\cite[(2.36)]{ewdon2}. A more detailed version of~\eqref{this} is the
  following:
  $$
  \D_X(e^{\Sigma+\lambda\nu})(m)=\sum_k 2^{-4d(k)}
  e^{\left(\frac{m}{2\pi}\right)\ip{x,\Sigma}}
  e^{\left(\frac{m}{2\pi}\right)^2\cdot 2\lambda} Z_k(m).
  $$
\end{rem}


\subsection{Abelian Localization of Seiberg--Witten theory}
\label{chswabloc}

Recall that the equivariant SW partition function is given by
$$
Z_c(m)=\int_{\calA_L\times_{\calG_L}\Gamma(W^+)}
e_{\calG}(\calE_1,s_1)\wedge e_{\calG}(\calE_2,s_2)\wedge\Gamma_{\calG},
$$
where
\begin{eqnarray*}
  \calE_1&=&\left(\calA_L\times\Gamma(W^+)\right)\times_{\calG_L}
  \Omega^2_+(X),\\
  \calE_2&=&\left(\calA_L\times\Gamma(W^+)\right)\times_{\calG_L}
  \Gamma(W^-).\\
\end{eqnarray*}
$S^1$ acts on $\calA_L\times\Gamma(W^+)$ by scalar multiplication
on the spinors $\Gamma(W^+)$, and we can consider the based gauge group
$\calG_L^0$ in the split exact sequence
$$
1\to\calG_L^0\to\calG_L\to S^1\to 1,
$$
with $\calG_L^0$ acting freely on $\calA_L\times\Gamma(W^+)$, and $S^1$
acting on $\calA_L\times\Gamma(W^+)/\calG_L^0$.  The fixed point set of the
$S^1$ action is:
$$
M_0=\calA_L/\calG_L^0.
$$
As in $\S$\ref{dwabloc}, we write the partition function in terms of
$S^1$-equivariant Euler classes and then formally apply the abelian
localization theorem.  Note that $\Gamma_{S^1}\big|_{M_0}=1$ as before.  We
get
\begin{equation*}
  SW(c,m)=\int_{\calA_L/\calG_L^0\times 0} 
  e(\calE_1,s_1)\wedge\frac{e_{S^1}(\calE_2,s_2)}
  {e_{S^1}(\nu_{\calA_L/\calG_L^0\subset\calA_L
      \times_{\calG_L^0}\Gamma(W^+)},0)},
\end{equation*} 
where the Euler classes are restricted to $\calA_L/\calG_L^0$. Moreover
notice that
\begin{eqnarray*}
  e_{S^1}(\calE_1,s_1)\big|_{\calA_L/\calG_L^0\times 0}&=&
  e(\calE_1,F_A^+)\big|_{\calA_L/\calG_L^0},\\
  e_{S^1}(\calE_2,s_2)\big|_{\calA_L/\calG_L^0\times 0}&=&
  e_{S^1}(\calE_2,0)\big|_{\calA_L/\calG_L^0}.
\end{eqnarray*}
We compute the following quotient of Euler classes as in the previous section
(see also~\cite[Prop. 69]{radu}). We have:
$$
\frac{e_{S^1}(\calE_2,0)}{e_{S^1}(\nu,0)}=
\left(\frac{m}{2\pi}\right)^{-\text{Ind}({\Dirac}_A)}
c_{\text{tot}}(-\ind({\Dirac}_A))\left(\frac{2\pi}{m}\right).
$$
Here
\begin{equation}
  \Ind({\Dirac}_A)=-\Ind(d^++d^*)+2s(c),
\end{equation}
where $2s(c)=\dim\calM_c$.  In the partition function case we have
$2s(c)=0$, thus,
\begin{equation}
  \label{diracindex}
  \Ind({\Dirac}_A)=-\Ind(d^++d^*)=\frac{\chi+\sigma}{2}.
\end{equation}
Then the expression for the equivariant SW partition function becomes
\begin{equation}
  \label{swabloc}
  Z_c(m)=\int_{\calA_L/\calG_L} e(\calA_L\times_{\calG_L}\Omega^2_+,F_A^+)
  \wedge\left(\frac{m}{2\pi}\right)^{-\frac{\chi+\sigma}{2}}\cdot\1 ,
\end{equation}
from the same reasons of dimensionality as in the Remark~\ref{dimaspect}.  
\begin{rem}
  \label{swrem}
  Note that, using physics arguments, one has to introduce a massive term in
  the MQ expression of the SW integral, which corresponds to twisting the
  $N=2$ SUSY Yang-Mills theory coupled with a massive hypermultiplet. For
  details, one can see e.g.~\cite{park},~\cite{ewmono}.  This comes down to a
  perturbation of the SW equations, equivalently one can consider
  \spinc--structures $\tilde{c}=c+2l$, where $l^2<0$ and
  $$
  c\cdot l = -\Ind\Dirac_A+\frac{\chi+\sigma}{2}=-2s(c).
  $$
  Furthermore, if one chooses a basis $\Sigma_1,\dots,\Sigma_{b_2}$ of
  $H_2(X)$, and then write $l$ in this basis, then there is a constant
  $\alpha$ such that
  $$
  (c+2l)\cdot\Sigma_i = c\cdot\Sigma_i + 2\alpha\Sigma_i^2.
  $$
  In the case when $X$ has simple type, the SW invariant for the perturbed
  \spinc--structure will equal the unperturbed invariant, or it will be zero.
  For details, one can see~\cite{szabo1}.  This remark will be very useful in
  $\S$\ref{conjecture}.
\end{rem}
Recall that the observables of SW theory are obtained by wedging with the
first Chern class of the line bundle $\calL\big|_{pt}$. Using the same
arguments for the localization of $\mu(\nu)$ in the Donaldson theory, we get:
\begin{equation}
  \label{swobsres}
  c_{1,\calG_L}(\calL\big|_{pt})\big|_{M_0}=-\left(\frac{m}{2\pi}\right)^4\cdot\1
\end{equation}
Then the formula for the equivariant SW invariant is:
\begin{eqnarray}
  \label{swobs1}
  SW(c,m)&=&\int_{\calA_L/\calG_L} e(\calA_L\times_{\calG_L}\Omega^2_+,F_A^+)
  \wedge\left(\frac{m}{2\pi}\right)^{-\frac{\chi+\sigma}{2}-2s}\cdot\1
  \wedge (-1)^{2s}\left(\frac{m}{2\pi}\right)^{4s}\cdot\1\nonumber\\
  &=& \int_{\calA_L/\calG_L} e(\calA_L\times_{\calG_L}\Omega^2_+,F_A^+)
  \wedge\left(\frac{m}{2\pi}\right)^{-\frac{\chi+\sigma}{2}+2s}\cdot\1\nonumber\\
  \,
\end{eqnarray}

\begin{rem}
  \label{swrem2}
  A heuristic comparison of~\eqref{swobs1} with the formula for the SW
  partition function~\eqref{swabloc} suggests one can, in some cases,
  perturb the \spinc--structure $c$, for which $2s(c)>0$, with a line bundle
  $l$, such that $(c+2l)^2-(2\chi+3\sigma)=8s(c+2l)=0$, in order to obtain a
  zero dimensional moduli space.  In this case, if we formally replace
  (modulo $2\pi$) $m$ with $m^{2s(c)}$, we can conjecture the following
  formula which relates the SW invariants for positive dimensional moduli
  spaces with SW invariants for zero dimensional moduli spaces:
  $$
  SW(c,m^{2s})= SW(c+2l,m)
  $$
  The above formula is an analog of the formula obtained in~\cite[Theorem
  1.3]{szabo1}, for $l=\Sigma$ of negative square, together with certain
  conditions on $\Sigma$. Note that in general there are obstructions for
  finding $l$, and this perturbation cannot be done in complete generality.
  Without assuming that the moduli space for the \spinc--structure $c+2l$ has
  dimension zero, we can still write:
  \begin{equation}
    \label{swstillcanwrite}
    SW(c,m)=\left(\frac{m}{2\pi}\right)^{2s(c)-2s(c+2l)} SW(c+2l,m)
  \end{equation}
\end{rem}


\section{Witten's Conjecture}
\label{conjecture}

In this section we compare the formulas previously obtained by abelian
localization of DW and SW theories, and give a formal derivation for Witten's
formula.  Witten's conjecture states the following:
\begin{conj}
  Let $X$ be a compact simply connected smooth 4-manifold with $b_2^+$ odd
  and greater than or equal to 3. Then $X$ has Donaldson simple type if and
  only if it has SW simple type. Moreover, if $X$ has simple type then the
  Kronheimer and Mrowka basic classes~\cite{km} agree with the SW basic
  classes and
  \begin{eqnarray}
    \label{xxx}
    \D_X\left(\left(1+\frac{u}{2}\right)e^{\Sigma}\right)&=&
    2^{2+\frac{7\chi+11\sigma}{4}}e^{\Sigma\cdot\Sigma/2}
    \sum_{\stackrel{c}{\text{basic classes}}}SW(c)\cdot e^{c\cdot\Sigma}\nonumber\\
    \D_X\left(\left(1-\frac{u}{2}\right)e^{\Sigma}\right)&=&
    2^{2+\frac{7\chi+11\sigma}{4}}i^{\frac{\chi+\sigma}{4}}e^{-\Sigma\cdot\Sigma/2}
    \sum_{\stackrel{c}{\text{basic classes}}}SW(c)\cdot e^{-ic\cdot\Sigma}
  \end{eqnarray}
  for $\Sigma\in H_2(X)$ and $u$ the generator of $H_0(X)$.  Note
  that~\eqref{xxx} is equivalent to~\eqref{conjintro} (see e.g.~\cite{km}).
\end{conj}

We do not prove that D-simple type is equivalent to SW-simple type. We only
give a derivation of a similar formula relating DW and SW invariants without
imposing the simple type conditions. If we assume both simple type
conditions, we obtain exactly Witten's formula~\eqref{conjintro}.  The
partition functions case was studied in~\cite{vajiac}.


\subsection{A direct derivation}

The derivation of the general conjecture relating the Donaldson generating
series with the SW invariants is as follows.  Recall that the $k$-th
equivariant Donaldson polynomial for observables $\calO=\Sigma^a\wedge\nu^b$
is given by~\eqref{dwobs1}, and the SW invariant for a \spinc--structure $c$
is given by~\eqref{swobs1}. 

In order to identify the integrals inside both formulas~\eqref{dwobs1}
and~\eqref{swobs1}, we must match up the formal dimension of the spaces
$\calA_{l}/\calG_{l}$, as defined in Remark~\ref{dimaspect}.  Using
Remark~\ref{dwrem}, we write the Donaldson generating series as a formal sum
of the partition functions $Z_k(m)$. The integral obtained will have formal
dimension zero, with respect to the Donaldson moduli space.  

However, $x^2=l_x^2$ does not equal $2\chi+3\sigma$, so we cannot identify
$x$'s with \spinc--structures for the SW partition function. Thus we first
identify the DW generating series with SW invariants for positive dimension
moduli spaces, and we obtain a general formula relating the invariants,
without assuming any simple type conditions.  Then, assuming both simple
types for our manifold $X$, the general formula specializes to Witten's
conjectured formula~\eqref{conjintro}.

Using~\eqref{dwobs1}, we write the Donaldson generating
series~\eqref{dwgeneratingfunction} as follows:
\begin{eqnarray}
  \label{bebe}
  \D_X(e^{\Sigma+\lambda\nu})(m) &=&\sum_d\sum_{a+2b=2d}\frac{(2\lambda)^{b}}{a!b!}
  \sum_{x^2=-4k} 2^{-(\chi+\sigma)-4d}\ip{x,\Sigma}^{a}
  \left(\frac{m}{2\pi}\right)^{2d-2s}\nonumber\\
  &&\qquad\qquad\cdot\int_{\calA_{l_x}/\calG_{l_x}} 
  e(\calA_{l_x}\times_{\calG_{l_x}}\Omega^2_+,F_A^+)\wedge
  \left(\frac{m}{2\pi}\right)^{-\frac{\chi+\sigma}{2}+2s}\cdot\1\nonumber\\
  \,
\end{eqnarray}
Here $2d=2d(k)=a+2b$ is the dimension of the Donaldson moduli space
$\calM_k$, and $2s=2s(c)$ is the dimension of the Seiberg--Witten moduli
space $\calM_c$.

A useful approach is to consider perturbations of the line bundles $l_x$ for
the integral in~\eqref{bebe}, as this integral is similar to the integral
in~\eqref{swobs1}.  This leads us to replace the cohomology classes $x=2y$
above, which have square $x^2=-4k$, with classes of the form $l+2y$, for a
fixed \spinc--structure $l$, as mentioned in Remark~\ref{swrem}. Here
$l^2\equiv 0\mod 2$.  Using a blowup/blowdown argument (see
e.g.~\cite{fs1},~\cite{fs2}), we claim that we actually can take $l^2=0$.
Therefore, by Remark~\ref{swrem}, the integral in~\eqref{bebe} can be
identified with $SW(y,m)$ in~\eqref{swobs1}.  Under the change of variables
$x=l+2y$, the sum $\displaystyle\sum_{x^2=-4k}$ in~\eqref{bebe} becomes
$\displaystyle\sum_{y^2=-k}$.  Moreover, the factor $\ip{x,\Sigma}$ becomes
$2\cdot\left(\ip{y,\Sigma}+\frac{1}{2}\Sigma^2\right)$ (see again
Remark~\ref{swrem}). Thus~\eqref{bebe} becomes
\begin{eqnarray}
  \label{bebe1}
  \hspace{-1cm}\D_X(e^{\Sigma+\lambda\nu})(m) &=&
  \sum_d\sum_{a+2b=2d}\frac{(2\lambda)^{b}}{a!b!}
  \sum_{y^2=-k} 2^{-(\chi+\sigma)-4d+a}
  \left\langle y,\Sigma + \frac{1}{2}\Sigma^2\right\rangle^{a}\cdot\nonumber\\
  &&\qquad\qquad\qquad\qquad\qquad\cdot
  \left(\frac{m}{2\pi}\right)^{2d-2s} SW(y,m).
\end{eqnarray}
As in Remark~\ref{thetwofactor}, we now adopt the topological convention that
$k$ is positive, which multiplies the right hand side of~\eqref{bebe1} by
two. Replacing $y$ by $-iy$, and $m$ by $-m$, we can include also
\spinc--structures $y$ of positive square, so we can rewrite~\eqref{bebe1} as
a sum over all \spinc--structures, i.e. over all classes $c\in H^2(X)$.
Specifically, we obtain
\begin{eqnarray}
  \label{bebe2}
  &&\hspace{-1cm}\D_X\left(e^{\Sigma+\lambda\nu}\right)(m) = 
  \sum_{c\in H^2(X)} 2^{1-(\chi+\sigma)-4d+a}
  \left(\frac{2\pi}{m}\right)^{2s(c)}
  SW(c,m)\cdot e^{\left(\frac{m}{2\pi}\right)(c\cdot\Sigma+\frac{1}{2}\Sigma^2)}
  \cdot e^{\left(\frac{m}{2\pi}\right)^2 (2\lambda)}\nonumber\\
  &&\qquad + i^{\frac{\chi+\sigma}{4}}
  \sum_{c\in H^2(X)} 2^{1-(\chi+\sigma)-4d+a}
  \left(\frac{2\pi}{m}\right)^{2s(c)}
  SW(c,m)\cdot e^{\left(\frac{m}{2\pi}\right)(-ic\cdot\Sigma-\frac{1}{2}\Sigma^2)}
  \cdot e^{\left(\frac{m}{2\pi}\right)^2 (-2\lambda)}\nonumber\\
  \,
\end{eqnarray}
In~\eqref{bebe2} we have used
$$
SW(-ic,-m)=i^{\frac{\chi+\sigma}{4}}\cdot SW(c,m).
$$
The degree $2b$ of the observable $\mu(\nu)$ from Donaldson theory can be
identified with the dimension $2s$ of the Seiberg--Witten moduli space, so
$a+2s=2d$.  Thus
\begin{equation}
  \label{bebe3}
  2^{1-(\chi+\sigma) -4d(k) + a}=2^{1-(\chi+\sigma) -2d(k) -2s(l+2y)}.
\end{equation}
Furthermore, by Proposition~\ref{propdimD} and~\eqref{propdimSW}, one easily
checks that
$$
2d(k)= -2s(l+2y) -8s(y) - (\chi+\sigma) - \frac{7\chi+11\sigma}{4}.
$$
Therefore~\eqref{bebe3} equals 
\begin{equation}
  \label{dimsequal}
  2^{8s(y) + 1+ \frac{7\chi+11\sigma}{4}}.
\end{equation}
Combining~\eqref{bebe2},~\eqref{bebe3}, and~\eqref{dimsequal} lead us to the
following conjectured formula, which expresses the Donaldson generating
series in terms of Seiberg--Witten invariants. Note that the formula below
does not assume any of the simple type conditions.
\begin{conj}   For $X$ a compact simply connected smooth 4-manifold with $b_2^+$ odd,
  and $b_2^+\ge 3$, the equivariant DW generating series can be expressed via
  equivariant SW invariants as follows:
  \begin{eqnarray}
    \label{preformula}
    &&\D_X\left(e^{\Sigma+\lambda\nu}\right)(m) = 2^{1+\frac{7\chi+11\sigma}{4}}
    \sum_c 2^{8s(c)}\left(\frac{2\pi}{m}\right)^{2s(c)}
    SW(c,m)\cdot e^{\left(\frac{m}{2\pi}\right)(c\cdot\Sigma+\frac{1}{2}\Sigma^2)}
    \cdot e^{\left(\frac{m}{2\pi}\right)^2 (2\lambda)}\nonumber\\
    &&\qquad +2^{1+\frac{7\chi+11\sigma}{4}} i^{\frac{\chi+\sigma}{4}}
    \sum_c 2^{8s(c)}\left(\frac{2\pi}{m}\right)^{2s(c)}
    SW(c,m)\cdot e^{\left(\frac{m}{2\pi}\right)(-ic\cdot\Sigma-\frac{1}{2}\Sigma^2)}
    \cdot e^{\left(\frac{m}{2\pi}\right)^2 (-2\lambda)}\nonumber\\
    \,
  \end{eqnarray}
\end{conj}
The sums in~\eqref{preformula} run over all $c\in H^2(X)$.  For these sums to
be finite, we must assume e.g. that $X$ has finite type in the sense
of~\cite{km3} or~\cite{munoz}. In this case, identifying the degree zero
components in~\eqref{preformula} (the expression being regular at $m=0$), one
obtains a formula for the Donaldson generating function for manifolds of
finite type, formula that looks similar with the ones proved in~\cite{munoz},
or described in~\cite{km3}.
\begin{rem}
  It is important to mention here also that, because of lack of examples,
  \eqref{preformula} cannot be checked for accuracy.
\end{rem}

If $X$ has D-simple type, we replace $d$ by $d'$, where
$$
d'\equiv\frac{1}{4}(\chi+\sigma)\mod 2.
$$
This allows us to consider $k\in\frac{1}{2}\Z$. A direct computation,
starting from~\eqref{bebe}, shows that the D-simple type condition reduces
$\displaystyle\sum_{a+2b=2d}$ in~\eqref{bebe} to the cases $b=0$ or $b=1$;
the $b=1$ case can be reduced further to $s(c)=0$, by dimensional arguments
and relations between Donaldson invariants.  This simplifies the computation
above, in the sense that the integrals inside the DW generating
function~\eqref{bebe} can be directly identified with SW partition
functions~\eqref{swobs1}.  Moreover, assuming also that $X$ has SW simple
type, the terms $\displaystyle\sum_c$ in~\eqref{preformula} are finite sums
over all SW basic classes.  In this case, we obtain:
\begin{eqnarray}
  \label{Wequivfinalformula}
  \D_X\left(e^{\Sigma+\lambda\nu}\right)(m) &=& 2^{1+\frac{7\chi+11\sigma}{4}}
  \sum_{c\hbox{ \tiny{basic}}} 
  SW(c,m)\cdot e^{\left(\frac{m}{2\pi}\right)(c\cdot\Sigma+\frac{1}{2}\Sigma^2)}
  \cdot e^{\left(\frac{m}{2\pi}\right)^2 (2\lambda)}\nonumber\\
  &+& 2^{1+\frac{7\chi+11\sigma}{4}} i^{\frac{\chi+\sigma}{4}}
  \sum_{c\hbox{ \tiny{basic}}} 
  SW(c,m)\cdot e^{\left(\frac{m}{2\pi}\right)(-ic\cdot\Sigma-\frac{1}{2}\Sigma^2)}
  \cdot e^{\left(\frac{m}{2\pi}\right)^2 (-2\lambda)}\nonumber\\
  \,
\end{eqnarray}
Identifying the degree zero components in~\eqref{Wequivfinalformula}, we
derive Witten's conjectured formula:
\begin{eqnarray*}
  \label{Wfinalformula}
  \D_X\left(e^{\Sigma+\lambda\nu}\right)&=&\\
  2^{\scriptstyle{1+\frac{7\chi+11\sigma}{4}}}
  &\cdot&\left[e^{\left(\frac{1}{2}\Sigma^2+2\lambda\right)}
    \sum_c SW(c)\cdot e^{\scriptstyle{c\cdot\Sigma}}+i^{\frac{\chi+\sigma}{4}}
    e^{\left(-\frac{1}{2}\Sigma^2-2\lambda\right)}
    \sum_c SW(c)\cdot e^{\scriptstyle{-ic\cdot\Sigma}}\right].\\
\end{eqnarray*}

Note that the factor $\ex{\Sigma^2/2}$ in Witten's formula usually appears
via the polarization identity for K3 surfaces (see~\cite[p.  690]{km}), or by
performing Gaussian integrals on the functional
integral~\cite{park},\cite{ewdon2},\cite{ewmono}. In our computation it comes
from perturbations of \spinc--structures.


\section*{Acknowledgments}

The author is extremely grateful to his former advisor, Prof. S. Rosenberg,
for all the guidance and support in writing this paper. Moreover, the author
would like to thank Profs. R. Constantinescu, P.  Feehan, D. Freed, T.
Kimura, D. Ruberman, J. Weitsman, for all their useful comments and
suggestions.



\begin{thebibliography}{999}
  

\bibitem{ab} M.F. Atiyah, R. Bott, {\em The moment map and equivariant
    cohomology}, Topology {\bf 23} (1984), 1--28.
  
\bibitem{aj} M.F. Atiyah, L. Jeffrey, {\em Topological Lagrangians and
    cohomology}, J. Geom. Phys. {\bf 7} (1990), 119--136.
  
\bibitem{austin} D. Austin, P. Braam, {\em Equivariant homology}, Math. Proc.
  Cambridge Philos. Soc. {\bf 118} (1995), 125--139.
  
\bibitem{axsinger} S. Axelrod, I. M. Singer, {\em Chern-Simons perturbation
    theory}, Proc. XXth Int. Conf. Diff. Geom. Methods in Theor. Phys. {\bf
    1,2} (1991), 3--45; {\em Chern-Simons perturbation theory II}, J. Diff.
  Geom. {\bf 39} (1994) 173--213.
  
\bibitem{bgv} N. Berline, E. Getzler and M. Vergne, {\em Heat Kernels and
    Dirac Operators}, Springer Verlag (1992).
  
\bibitem{blau} M. Blau, {\em The Mathai-Quillen formalism and topological
    field theory}, J. Geom. Phys. {\bf 11} (1993) 95--127.
  
\bibitem{radu} R. Constantinescu, {\em Circular Symmetry in Topological
    Quantum Field Theory and the Topology of the Index Bundle}, Ph.D. thesis,
  MIT, 1997.
  
\bibitem{cordes} S. Cordes, G. Moore, S. Ramgoolam, {\em Lectures on $2$D
    Yang-Mills theory, equivariant cohomology and topological field
    theories}, G\`eom\`etries fluctuantes en m\`ecanique statistique et en
  th\`eorie des champs (Les Houches, 1994), North-Holland, Amsterdam, 1996,
  505--682.
  
\bibitem{dk} S. Donaldson, P. Kronheimer, {\em The Geometry of
    Four-manifolds}, Oxford University Press, Oxford (1990).
  
\bibitem{feehan} P. M. N. Feehan, T.G. Leness, {\em $PU(2)$ monopoles, I:
    Regularity, Uhlenbeck compactness and transversality}, J. Diff. Geom.
  {\bf 49} (1998), 265--410; {\em $PU(2)$ monopoles, II: Highest-level
    singularities and relations between four-manifold invariants}, preprint,
  \texttt{dg-ga/9712005}; {\em $PU(2)$ monopoles and relations between
    four-manifold invariants}, Topology Appl. {\bf 88} (1998), 111--145.
  
\bibitem{fs1} R. Fintushel, R. J. Stern, {\em Rational blowdowns of smooth $4$-manifolds},
  J. Diff. Geom. {\bf 46} (1997), 181--235. 

\bibitem{fs2} R. Fintushel, R. J. Stern, {\em The blowup formula for Donaldson invariants},
  Ann. of Math. (2) {\bf 143} (1996), 529--546.

\bibitem{fu} D. Freed, K. Uhlenbeck, {\em Instantons and Four manifolds},
  Springer-Verlag New-York Inc. (1991).
  
\bibitem{givental} A. B. Givental, {\em Equivariant Gromov--Witten
    Invariants}, IMRN {\bf 13} (1996), 613--663.
  
\bibitem{guillemin} V. Guillemin, S. Sternberg, {\em Supersymmetry and
    equivariant de Rham theory}, Mathematics Past and Present,
  Springer-Verlag, Berlin (1999).
  
\bibitem{park} S. Hyun, J-S. Park, {\em $N=2$ topological Yang-Mills theories
    and Donaldson's polynomials}, J. Geom. Phys. {\bf 20} (1996), 31--53.
  
\bibitem{km} P. Kronheimer, T. Mrowka, {\em Embedded surfaces and the
    structure of Donaldson's polynomial invariants}, J. Diff. Geom. {\bf 43}
  (1995), 573--734.
  
\bibitem{km1} P. B. Kronheimer, T. S. Mrowka, {\em The genus of embedded
    surfaces in the projective plane}, Math. Res. Lett. {\bf 1} (1994),
  797--808.
  
\bibitem{km3} P. B. Kronheimer, T. S. Mrowka, {\em The structure of
    Donaldson's invariants for four-manifolds not of simple type}, preprint.
  
\bibitem{labastida_marino1} J. M. F. Labastida, M. Mari\~no, {\em A topological
    Lagrangian for monopoles on four-manifolds}, Phys. Lett. B {\bf 351}
  (1995), 146--152.
  
\bibitem{labastida_marino3} J. M. F. Labastida, M. Mari\~no, {\em Non-abelian
    monopoles and four-manifolds}, Nucl. Phys. B {\bf 448} (1995),
  373--395.
  
\bibitem{labastida_marino4} J. M. F. Labastida, M. Mari\~no, {\em Polynomial
    invariants for $SU(2)$ monopoles}, Nucl. Phys. B {\bf 456} (1995),
  633--668.
  
\bibitem{labastida_marino2} J. M. F. Labastida, M. Mari\~no, {\em Twisted N=2
    supersymmetry with central charge and equivariant cohomology}, Commun.
  Math. Phys {\bf 185} (1997), 37--71.
  
\bibitem{lly} B. H. Lian, K. Liu, S. T. Yau, {\em Mirror Principle I},
  Asian J. Math. {\bf 1} (1997), 729--763; {\em Mirror Principle II},
  preprint, \texttt{math.AG/9905006}; {\em Mirror Principle III}, preprint,
  \texttt{math.AG/9912038};
  
\bibitem{marino} M. Mari\~no, {\em The geometry of SUSY gauge theories in four
    dimensions}, Ph.D. thesis, \texttt{hep-th/9701128}.
  
\bibitem{mq} V. Mathai and D. Quillen, {\em Superconnections, Thom classes
    and equivariant differential forms}, Topology {\bf 25} (1986), 85--110.
  
\bibitem{uplane} G. Moore, E. Witten, {\em Integration over the $u$-plane in
    Donaldson theory}, Adv. Theor. Math. Phys. {\bf 1} (1997), 298--387.
  
\bibitem{munoz} V. Mu\~noz, {\em Donaldson invariants of non-simple type
    4-manifolds}, preprint, \texttt{math.DG/9909165}.
  
\bibitem{pt} V. Pistrig\'ach, A. Tyurin, {\em Localization of the Donaldson
    invariants along Seiberg--Witten classes}, preprint,
  \texttt{math.DG/9507004}.
  
\bibitem{ruantian} Y. Ruan, G. Tian, {\em Higher genus symplectic invariants
    and sigma models coupled with gravity}, Invent. Math. {\bf 130}, (1997),
  455--516.
  
\bibitem{salomon} D. Salamon, {\em Spin Geometry and Seiberg-Witten
    Invariants}, preprint, 1996.
  
\bibitem{nsew} N. Seiberg and E. Witten, {\em Electro-magnetic duality,
    monopole condensation, and confinement in $N=2$ supersymmetric Yang-Mills
    theory}, Nucl. Phys. B {\bf 426} (1994) 19--52; (Erratum Nucl. Phys. B
  {\bf 430} (1994) 485--486); {\em Monopoles, duality and chiral symmetry
    breaking in $N=2$ supersymmetric QCD}, Nucl. Phys. B {\bf 431} (1994),
  484--550.
  
\bibitem{szabo1} P. Oszv\'atz, Z. Szab\'o, {\em The symplectic Thom conjecture},
  preprint, \texttt{math.DG/9811087}.
  
\bibitem{szabo2} P. Oszv\'atz, Z. Szab\'o, {\em Higher type adjunction
    inequalities in Seiberg--Witten theory}, preprint.
  
\bibitem{taubes} C. H. Taubes, {\em The Seiberg-Witten invariants and
    symplectic forms}, Math. Res. Lett. {\bf 1} (1994) 809-822; {\em ${\rm
      SW}\Rightarrow{\rm Gr}$: from the Seiberg-Witten equations to
    pseudo-holomorphic curves}, J. Amer. Math. Soc. {\bf 9}, 3 (1996),
  845--918.
  
\bibitem{vajiac} A. Vajiac, {\em Localization Techniques in Topological
    Quantum Field Theories}, PhD. thesis, May 1999.

\bibitem{ewdon} E. Witten, {\em Topological quantum field theory},
  Commun. Math. Phys. {\bf 117} (1988), 353--386.
  
\bibitem{ewdon2} E. Witten, {\em Supersymmetric Yang-Mills theory on a
    four-manifold}, J. Math. Phys. {\bf 35} (1994), 5101--5135.
  
\bibitem{ewmono} E. Witten, {\em Monopoles and four-manifolds}, Math. Res.
  Lett. {\bf 1} (1994), 769--796.
  
\bibitem{ewsig} E. Witten, {\em Topological sigma models}, Commun. Math.
  Phys. {\bf 118} (1988), 411--449.
  
\bibitem{wu} S. Wu, {\em On the Mathai-Quillen formalism of topological sigma
    models}, J. Geom. Phys. {\bf 17}, 4 (1995), 299--309.

\end{thebibliography}
\end{document}